\documentclass[%
 aip,
 amsmath,amssymb,
 reprint,%
]{revtex4-1}

\usepackage{graphicx}
\usepackage{dcolumn}
\usepackage{bm}
\usepackage{subfig}
\usepackage{booktabs}
\usepackage[flushleft]{threeparttable}

\usepackage[utf8]{inputenc}
\usepackage[T1]{fontenc}
\usepackage{mathptmx}
\usepackage{multirow}
\begin{document}

\preprint{AIP/123-QED}

\title[Hydrodynamics of a Compound Drop in Plane Poiseuille Flow]{Hydrodynamics of a Compound Drop in Plane Poiseuille Flow}

\author{Vignesh Thammanna Gurumurthy}
 \email{vigneshtg@gmail.com}
\author{S. Pushpavanam}%
 \email{spush@iitm.ac.in}
\affiliation{ 
Department of Chemical Engineering, Indian Institute of
  Technology Madras, Chennai-600036, India
}%

\date{\today}

\begin{abstract}
  We numerically investigate the hydrodynamics of a two-dimensional compound
  drop in a plane Poiseuille flow under Stokes regime. A neutrally buoyant,
  initially concentric compound drop is released into a fully developed flow,
  where it migrates to its equilibrium position. Based on the results, we find
  that the core-shell interaction affects the dynamics of both the core and the
  compound drop. During the initial transient period, the core revolves about
  the center of the compound drop due to the internal circulation inside the
  shell. At equilibrium, depending upon the nature of the flow field inside the
  shell, we identify two distinct core behaviors: stable state and limit-cycle
  state. In the stable state, the core stops revolving and moves outward very
  slowly. The core in the limit-cycle state continues to revolve in a nearly
  fixed orbit with no further inward motion. The presence of core affects both
  deformation and migration dynamics of the compound drop. A comparison with the
  simple drop reveals that the core enhances the deformation of the compound
  drop. The outward moving core in stable state pushes the compound drop towards
  the walls, while the revolving core in limit-cycle state makes the compound
  drop to oscillate at its equilibrium position. The migration of the compound
  drop also affects the eccentricity of the core significantly. From the
  parametric study, we find that the core affects the compound drop dynamics
  only at intermediate sizes, and increase in any parameter sufficiently causes
  a transition from limit-cycle state to stable state.
\end{abstract}

\maketitle

\section{Introduction}
A compound drop, also called a double emulsion, is a multicomponent liquid system
consisting of one or more drops encapsulated inside another immiscible drop. The
outermost drop is often referred as the shell and the inner ones as the core.
The shell functions as a protective layer to the core from the ambient fluid.
This feature makes the compound drops to be used as a delivery system, where the
core can be loaded with chemical reactants \cite{Liu2016}, drugs
\cite{C7LC00242D,C7CS00263G}, and food additives \cite{Muschiolik2017}. In addition, it is
also suited for in-situ culturing of cells and bacteria
\cite{C7CS00263G,Hati2016,Alkayyali2019}. Thus compound drops are highly desirable for
applications related to bioanalysis, cosmetics, food and pharmaceutical
industries.

Majority of the fundamental works on compound drops have focussed primarily on
their production techniques and their dynamics in three basic flow geometries:
(i) translation in a quiescent liquid
\cite{Johnson1985,Sadhal1985,Bazhlekov1995,Homma2011}, dynamics in (ii)
extensional flows \cite{StoneHA1990,Wang2016,Wang2018} and (iii) linear shear
flows
\cite{StoneHA1990,Smith2004,qu2012,Hua2014,Chen2015,Chen2015a,Vu2019a}. But, the
pressure-driven flow (also known as Poiseuille flow) often encountered in
microfluidic and biological systems has not been examined thoroughly yet. The
stability of the compound drops moving in these flows play a huge role in
applications related to targeted release of active ingredients in the core
\cite{C7LC00242D}. Thus, understanding the dynamics of compound drops in
Poiseuille flow is crucial for the aforementioned applications.

Very few studies have investigated the dynamics of compound drops in Poiseuille
flow \cite{Zhou2008,Song2010,Tao2013,Borthakur2018,Che2018}, but they are not
applicable at all conditions due to their underlying assumptions. The
theoretical work by Song \textit{et al.} \cite{Song2010} studied the dynamics
under Stokes regime in a capillary tube assuming both the shell and the core are
perfectly spherical. Their analysis shows that the drag force on the core and
the shell depends on various parameters such as the viscosity ratio between the
liquids (core-shell, shell-carrier), size ratio between the core and the shell,
and eccentricity of the core. The use of perfectly spherical drop assumption
limits their applicability since it is well known that the drops deform due to
the confinement effects \cite{GUIDO201089} which alters its shape. The numerical
works by Zhou \textit{et.al}\cite{Zhou2008} and Tau \textit{et.al}
\cite{Tao2013} investigated the morphological evolution of the compound drop
moving in circular tube with a gradual contraction using phase field and
boundary element methods, respectively. Both studies show that the core affects
the deformation of the compound drop and also prolongs the transit time in the
contraction zone. Studies by Che \textit{et.al}\cite{Che2018} and Borthakur
\textit{et.al} \cite{Borthakur2018} simulated the dynamics of compound drop in
circular microchannels under axisymmetric conditions using level-set and
volume-of-fluid methods respectively. They investigated the temporal evolution
of the drop shape, velocity fields, and the eccentricity of the core under the
influence of parameters such as size of the compound drop, radius ratio,
viscosity ratio and the capillary number.

The use of axisymmetry in the above studies limits their results to compound
drops moving in circular microchannels. In addition, the axially moving drops
(core and shell) are always at the centerline of the channel due to
axisymmetry. But we know from theory of migration for simple drops
\cite{Chan1979}, that only large drops comparable to channel size occupies the
center, whereas smaller drops often migrate to an off-centered position. Recent
experiments \cite{Yu2019} have shown that the migration of compound drops in
flow focusing geometries experience enhanced deformation. Thus, the effect of
migration of the compound drop on the core, and the effect of core on the
migration of the compound drop has not yet been understood clearly, and this
forms the primary focus of this work.

In this work, we investigate the dynamics of the compound drop in a plane
Poiseuille flow, which includes the migration and deformation dynamics of the
core and the compound drop. The objectives of this study are to understand the
(i) effect of migration of the compound drop on the dynamics of the core, (ii)
effect of core on the migration and deformation of the compound drop, and (iii)
analyze their dynamics under the influence of different parameters
involved.

Rest of the paper is organized as follows. In section II, we define the problem,
list the governing equations and describe the numerical method adopted. In
section III, we first describe the dynamics of the core followed by the compound
drop, and finally the influence of different parameters on the
equilibrium behaviors of the core and the compound drop. Finally, we summarize
the results in section IV.
\section{Problem formulation}
\subsection{Problem definition}

\begin{figure}
\centering
    \includegraphics[scale=0.7]{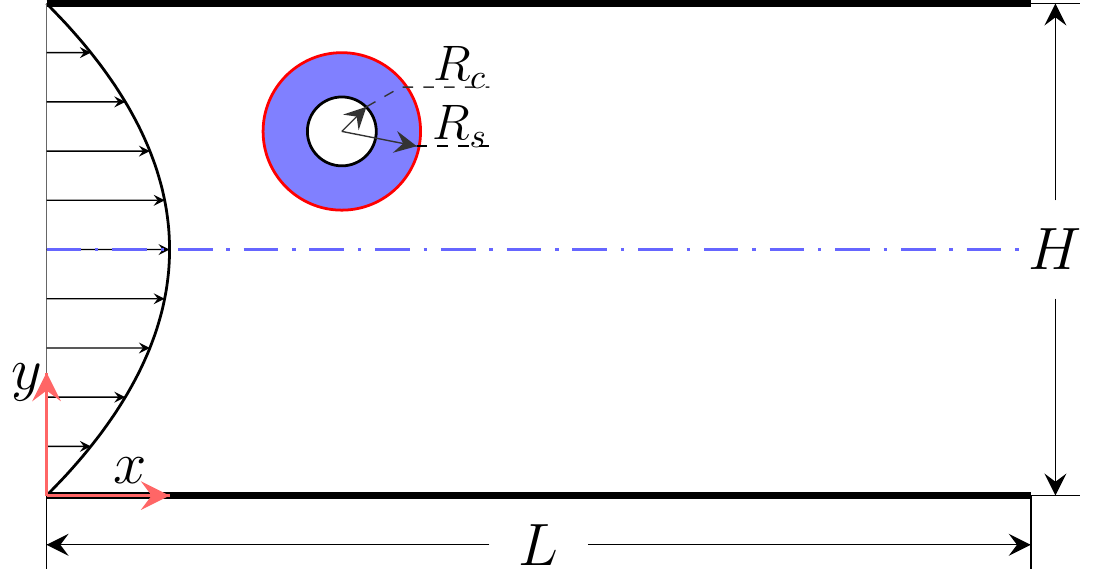}
\caption{Schematic of the computational domain. A neutrally buoyant, initially concentric compound drop is released into the fully developed flow.}
  \label{fig_1}
\end{figure}

Consider a concentric compound drop released in a fully developed plane
Poiseuille flow as shown in Fig. \ref{fig_1}. We choose the subscripts $c$, $s$,
and $d$ to denote the parameters related to the core, shell and the compound
drop, respectively. The radius of the undeformed core and shell are denoted as
$R_c$ and $R_s$, respectively. We choose the commonly used core-shell-carrier
fluid configuration in experiments: water-in-oil-in-water (W/O/W). This choice
simplifies our problem to a two-phase flow since the core and the carrier phase
are of the same liquid. The liquids are assumed to be Newtonian, and neutrally
buoyant.

\subsection{Governing equations}

We model the two-phase flow using Volume-of-Fluid (VOF) method, where a single
set of governing equations describe the flow in the domain occupied by the two
phases. The two phases are treated together as a continuum with variable
material properties which abruptly changes at the interface. Since the flow
rates associated with the microfluidic systems are typically low, Stokes flow
conditions are assumed. The governing equations are,

\begin{equation}
  \nabla \cdot \mathbf{U} = 0,
  \label{eq_mass}
  \end{equation}
  \begin{equation}
 - \nabla P + \nabla \cdot   \left(\mu\mathbf{D} \right) + \sigma \kappa \mathbf{n}
  \delta_s + \mathbf{S_p} =0,
  \label{eq_mom}
\end{equation}

where $\mu$, $\mathbf{U}$ and $P$ represent viscosity, velocity and pressure
field, respectively. The rate of strain tensor $\mathbf{D}$ is defined as
$(\nabla \mathbf{U} + \nabla \mathbf{U}^T)$. The third term on the right hand
side of Eq. (\ref{eq_mom}) represents the surface tension force acting at the
interface modelled as a body force term according to continuum surface force
formulation \cite{BRACKBILL1992}. The use of Dirac delta function $\delta_s$
ensures that this force acts only in the vicinity of the interface and zero
anywhere; Also, it acts along the direction of the normal $\mathbf{n}$ to the
interface. The symbols $\sigma$ and $\kappa$ denote the surface tension
coefficient and curvature of the interface, respectively. The flow through the
channel is driven by a constant pressure gradient which is introduced as a
source term $\mathbf{S_p}$ in Eq. (\ref{eq_mom}) (see appendix \ref{app_01} for
derivation). It is given by,
\begin{equation*}
  \mathbf{S_p} = \frac{8 \mu_c U_{m}}{H^2} \mathbf{\hat{x}},
\end{equation*}
where $U_{m}$ is the maximum velocity observed at the centerline of the channel
in the absence of the drop, and $H$ is the channel height.

The two phases are distinguished by the scalar field $f$, which represents
the volume fraction of the shell liquid. Computational cells with $f=1$ represent
the shell phase and $f=0$ represent either the core or the carrier phase.
Thus cells with  $f$ values between zero and one contain the interface whose
motion is governed by the advection equation,

\begin{equation}
  \frac{\partial f}{\partial t} + \nabla \cdot (f \mathbf{U})=0.
\label{eq_advec}
\end{equation}

The viscosity is now defined as,
\begin{equation}
  \mu = \mu_s f + (1-f) \mu_c.
\label{eq_properties}
\end{equation}

\subsection{Non-dimensionalization}

We non-dimensionalize the above equations (\ref{eq_mass})-(\ref{eq_properties})
using the channel height $H$, centerline velocity $U_m$, and the ratio $\tau=H/U_m$
as the characteristic length, velocity, and time scale, respectively. The
characteristic time $\tau$ scale used here is the time required for advecting the
volume fraction field i.e, interface. Pressure is scaled using the viscous
friction given by $\mu_cU_m/H$, and the viscosity in Eq. (\ref{eq_properties})
using $\mu_c$. We write the non-dimensionalized equations in the same form as
their dimensional counterparts for convenience, and they are given below,

\begin{subequations}
  \label{eq_nd}
  \begin{equation}
    {\nabla} \cdot {\mathbf{U}} = 0, \label{eq_con_nd} 
  \end{equation}
  \begin{equation}
- \nabla {P} + \nabla \cdot \left({\mu} \mathbf{{D}} \right) + \frac{{\kappa} \mathbf{n}\delta_s}{Ca}
+ 8 \mathbf{\hat{x}} = 0, \label{eq_navier_nd}
\end{equation}
\begin{equation}
  \frac{\partial f}{\partial {t}} + \nabla \cdot (f {\mathbf{U}})=0, \label{eq_adv_nd}
\end{equation}
\text{and the viscosity is given by,}
  \begin{equation}
    {\mu} = \mu_r f + (1-f). 
  \end{equation}
\end{subequations}

In Eq. (\ref{eq_nd}b), $Ca =\mu_c U_m/\sigma$, represents the capillary number
which is the ratio between viscous and capillary forces, and
$\mu_r=\mu_s/\mu_c$ in Eq. (\ref{eq_nd}d) represents the viscosity ratio
between the two liquids. In addition, the dynamics of the compound drop also
depends on the radius ratio $K =R_c/R_s$, and aspect ratio $\lambda=R_s/H$. The
radius ratio signifies the confinement effects of the shell on the core, while
the aspect ratio signifies the confinement effects the wall.

\subsection{Numerical method}
We solve the quasisteady Stokes equations using pseudotransient method. In this
method, by including the unsteady acceleration term in Eq. (\ref{eq_nd}b) we
construct an unsteady equation as shown below,
\begin{equation}
  \frac{\partial \mathbf{U}}{\partial t} = - \nabla {P} + \nabla \cdot \left({\mu} \mathbf{{D}} \right) + \frac{{\kappa} \mathbf{n}\delta_s}{Ca}
+ 8 \mathbf{\hat{x}}. \label{eq_nd_1}
  \end{equation}

  The transient solution of the above unsteady equation is marched in time until
  steady state is reached.

The governing equations are solved numerically using the open source code
Basilisk \cite{POPINET2003572,POPINET20095838} written in C programming
language. The equations are discretized using finite volume method where the
primitive variables ($u,~v, ~P, ~f$) are collocated at the cell centers. The
discretizations are second-order accurate in both space and time. Equations
(\ref{eq_nd}a) and (\ref{eq_nd_1}) are solved using a pressure projection
approach, and the advection equation (\ref{eq_adv_nd}) using an operator-split
algorithm. The surface tension force in Eq. (\ref{eq_nd_1}) is accurately
calculated using balanced force algorithm \cite{FRANCOIS2006141}, and the
curvature of the interface is computed from the volume fraction field using
height functions technique \cite{Afkhami2008,POPINET20095838}.

Basilisk provides an adaptive mesh refinement technique, which is used for
refining specific regions to improve accuracy. The refinement is based on a
wavelet transform of a given scalar field, which is used to assess its
discretization error. The grid is locally refined if the calculated error is
above the user-specified threshold and coarsened otherwise. In the simulations
presented here, the threshold values are fixed at $1\times 10^{-3}$ and
$1\times 10^{-2}$ for the two velocity components $(u,~v)$ and the volume
fraction field ($f$), respectively. These values ensure that the regions close
to the interfaces are sufficiently resolved for accurate calculation of the
surface tension force.

Due to the explicit treatment of surface tension force in basilisk, the time
step is calculated according to the Brackbill criterion \cite{BRACKBILL1992},

\begin{equation}
  \Delta t \leq \sqrt{\frac{\rho \Delta x^3}{\pi \sigma}}
\end{equation}

which is the time step required to resolve the propagation of shortest capillary
wave resolved by the mesh. By comparing equations (\ref{eq_mom}) and
(\ref{eq_nd_1}) the above condition reduces to,
\begin{equation}
  \Delta t \leq \sqrt{\frac{\Delta x^3 Ca}{\pi}}.
  \label{eq_dt}
\end{equation}
The above equation shows that the time step is directly proportional to the
capillary number i.e., deformation.

\subsection{Computational domain and boundary conditions}

As often reported in the literature \cite{Mortazavi2000,Stan2011,Lan2012},
simulating cross stream migration of a drop is time consuming because the rate
of migration is inversely proportional to the deformation of the drop. In
addition, the time steps given by Eq. (\ref{eq_dt}) decreases with the decrease
in $Ca$. Thus, it takes longer time for the drop to reach its equilibrium
location. Hence, we restrict our investigations to two dimensions. Nevertheless,
previous works \cite{Mortazavi2000,Stan2011} have shown that the 2D simulations
qualitatively captures the flow field obtained in the center plane of 3D
simulations. Also, there is a good quantitative agreement for the equilibrium
position of the drop calculated between the 2D simulations and the experiments
\cite{Mortazavi2000,Stan2011}.

The computational domain, as shown in Fig. \ref{fig_1}, is a rectangular
domain of size $L\times H$. No slip and no penetration boundary conditions are
enforced at the top and bottom walls, and a periodic boundary condition in the
horizontal direction. The size of the computational domain is chosen as $2
\times 1$ for smaller drops ($\lambda \leq 0.2$) and $4 \times 1$ for larger
drops. These different domain lengths for different drop sizes are chosen such
that the drop-drop interaction due to periodicity is negligible.

As mentioned above, since the migration of a drop to its equilibrium position is
time consuming, it is advisable to release the drops closer to its equilibrium
position. However, the equilibrium positions are not known a priori. Based on
the literature on migration of simple drops
\cite{Chan1979,Mortazavi2000,Stan2011}, and from few trial and error
simulations, we adopt the following approach. For simulations with $Ca < 0.25$
we release the concentric drop at $y_{i}=0.65$ and at $y_{i}=0.6$ for
$Ca \geq 0.5$.

\begin{figure}
  \subfloat[\label{fig_2a}]{
    \includegraphics[scale=1]{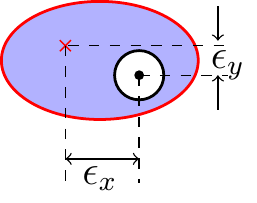}}
  \subfloat[\label{fig_2b}]{
    \includegraphics[scale=1]{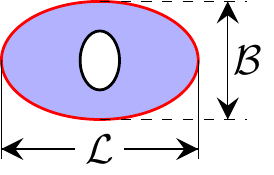}}
  \caption{Schematics showing the definition of (a) the eccentricity of the core
    and (b) the deformation of the compound drop. The black dot and the red
    cross markers in (2a) represent the centroid of the core and the compound
    drop, respectively.}
  \label{fig_2}
\end{figure}

We characterize the dynamics of the compound drop, and the core by studying
their migration and deformation behavior. The migration of the compound drop is
characterized by its height $y_d$ measured from the bottom wall. In case of the
core, we report its eccentricity which is defined as
$\vec{\epsilon}=\vec{x_c}-\vec{x_d}$ (see Fig. \ref{fig_2}(a)), where
$\vec{x_c}$, $\vec{x_d}$ are the centroids of the core and the compound drop,
respectively. The deformation characteristics of the core and the compound drop
are quantified by calculating the deformation parameter $\mathcal{D}$ proposed
by Taylor \cite{Taylor}, which is defined as
$\mathcal{D} = (\mathcal{L}-\mathcal{B})/(\mathcal{L}+\mathcal{B})$. Here,
$\mathcal{L}$ and $\mathcal{B}$ denote the length of the major and minor axis of
the deformed droplet as shown in Fig. \ref{fig_2}(b). Each simulation is stopped
shortly after equilibrium is reached i.e, when the parameters measured either
attain a constant value or enter a periodic state.

Based on grid convergence tests (see appendix \ref{app_1}), the size of the
smallest region resolved in our simulations is $1.95 \times 10^{-3}$. We also
have validated the code with the experiments and the simulations reported in
the literature on simple drops released in pressure-driven flows (see appendix
\ref{app_2}).

\section{Results and Discussions}
\subsection{Dynamics of the core}
In this section, we describe the characteristics of two distinct core behaviors
commonly observed in the simulations, and also explain its cause by analysing
the flow field inside the shell.

\subsubsection{Distinct core behaviors}
\label{sec_core}
\begin{figure*}
  \subfloat[]{
    \centering
    \includegraphics[scale=0.7]{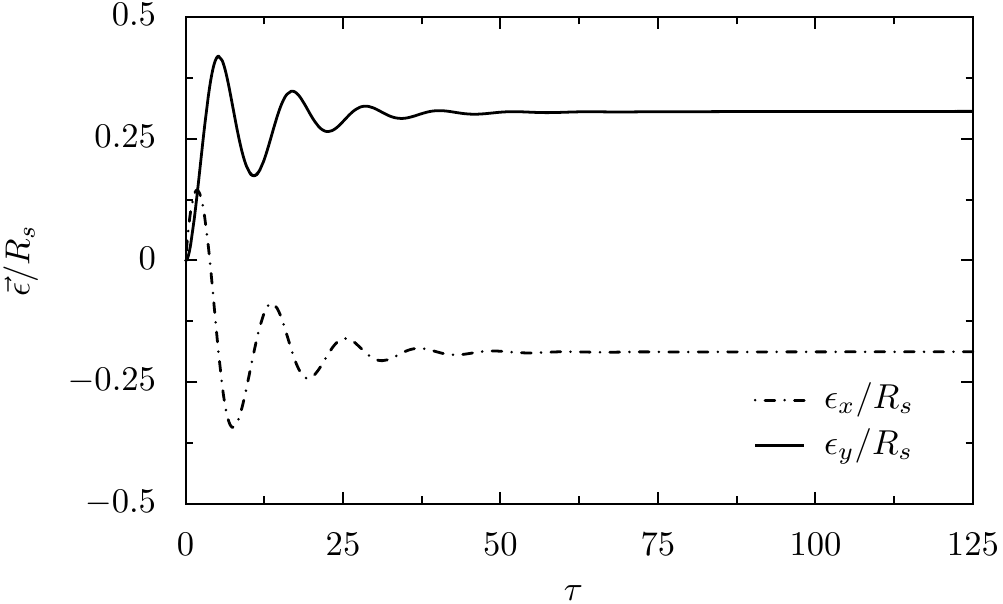}}
    \subfloat[]{
    \centering
    \includegraphics[scale=0.7]{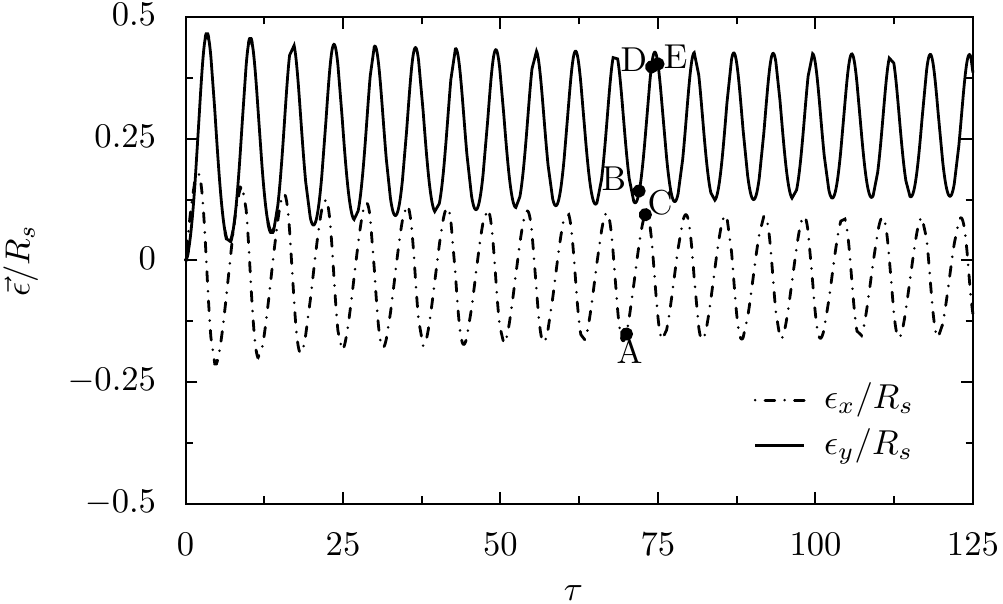}}

\subfloat[]{
    \centering
    \includegraphics[scale=0.7]{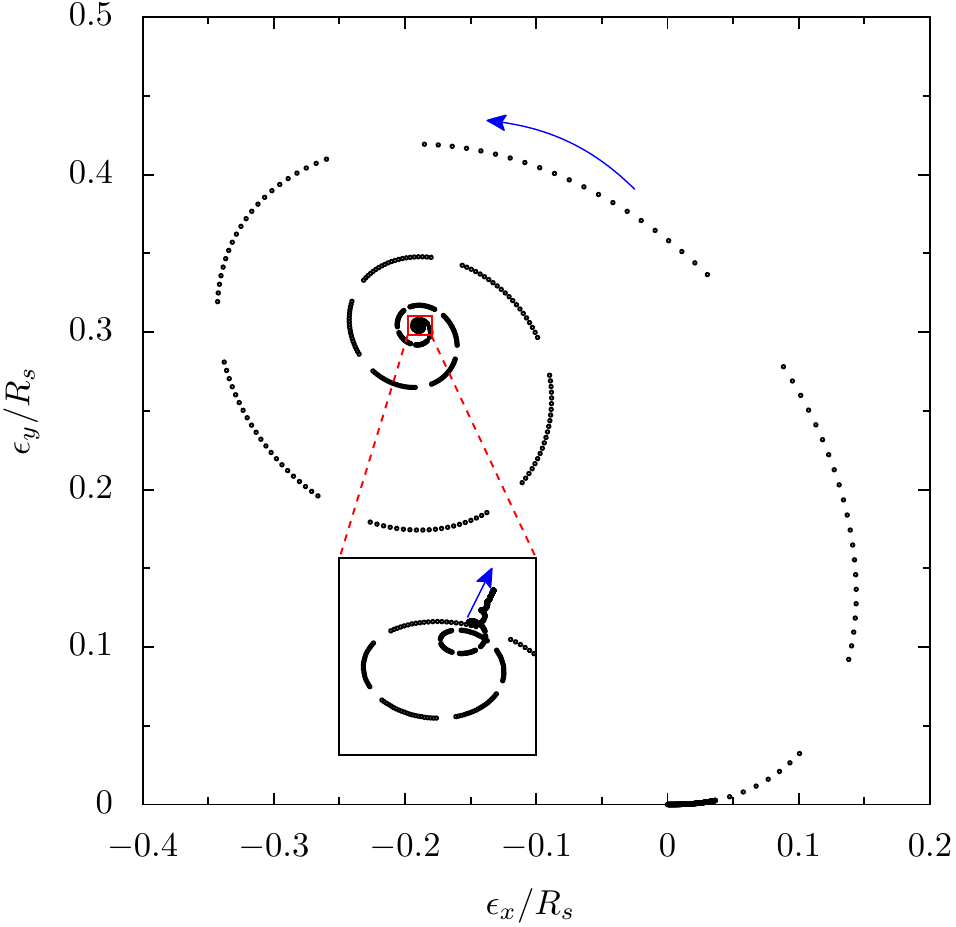}}
  \subfloat[]{
    \centering
    \includegraphics[scale=0.7]{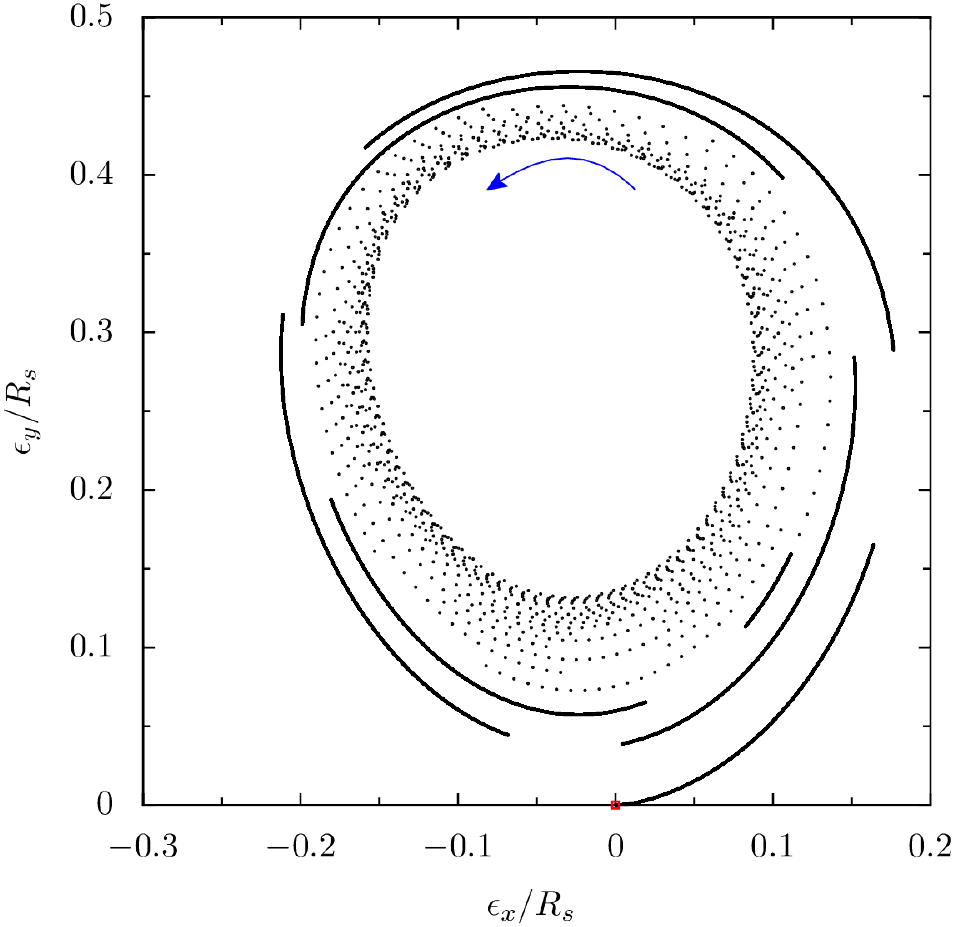}}  
  \caption{Temporal evolution of the eccentricity of the core under (a) stable
    state and (b) limit-cycle state. Trajectory of the core inside the shell
    under (c) stable state and (d) limit-cycle state. The gaps seen in the
    trajectory of the core are due to the data ignored because of periodic
    boundary conditions. The eccentricity and the trajectory of the core are
    measured relative to the centroid of the compound drop
    ($\vec{\epsilon}=\vec{x_c}-\vec{x_d}$). The common parameters between the
    two states are $\lambda=0.2$, $K=0.5$, $\mu_r=1$ and the capillary number
    was kept fixed at $Ca=1$ for the stable state and $Ca=0.05$ for the
    limit-cycle state. }
  \label{fig_3}
\end{figure*}

The temporal evolution of the eccentricity of the core is shown in Fig.
\ref{fig_3}(a) and Fig. \ref{fig_3}(b) under two different conditions: (i)
$Ca=1$ and (ii) $Ca=0.05$. The other parameters were kept fixed at $\lambda=0.2$
and $K=0.5$, for both the cases. Initially, under both conditions, we observe
oscillations in the eccentricity in both directions. Eventhough they are
slightly out of phase with each other, their magnitudes are nearly the same. In
the first case, the oscillations decay at a faster rate, eventually reaching a
constant value. We call this equilibrium behavior as stable state. The
oscillations in the second case decay slightly and enters a periodic state at
long-times where it continues to oscillate at constant amplitude. We refer this
periodic behavior as limit-cycle state. Thus at equilibrium, we observe two
distinct behaviors, which have also been observed for compound drops in linear
shear flows \cite{Chen2015,Chen2015a,Kim2017}.

The oscillations of the core observed in both directions indicate that it
revolves inside the shell around the centroid of the compound drop, as
illustrated in Fig. \ref{fig_3}(c) and Fig. \ref{fig_3}(d). The core, as it
revolves also moves inward in both cases. This revolving motion is due to the
internal circulation inside the shell \cite{Chen2011}, which is the result of
shearing of the shell interface due to the difference in velocities between the
interface and the carrier fluid. The anticlockwise direction of both the
circulation (see Fig. \ref{fig_4}(b) \& Fig. \ref{fig_4}(d)) and the revolving
motion of the core confirms that the core revolves due to the circulation.

In the stable state, as the core revolves, the length of its orbit reduces
rapidly in each cycle in the same fashion similar to the decay of oscillations
observed in Fig. \ref{fig_3}(a). As shown in the inset in Fig. \ref{fig_3}(c),
at equilibrium, the core stops revolving and starts translating outwards. The
velocity of the core in this outward motion can be inferred from the small
distance travelled to be very small. The core in the limit-cycle state, also
starts revolving and moving inward initially. But at equilibrium, it enters into
an egg shaped orbit (see Fig. \ref{fig_2}(d)), where its length in each cycle
remains nearly the same thereafter, and also, it does not move inward any
further.

\subsubsection{Underlying mechanism}
The two distinct core behaviors at equilibrium can be explained by analysing the
rotational flow field they experience inside the shell. The gradient of the
velocity field in any incompressible flow can be decomposed into strain rate
tensor and vorticity tensor. The former causes deformation, while the
latter causes rigid body rotation. A drop in pure straining field will undergo
planar extension: stretching and compressing the drop along the principal
axes. In a pure vorticity field, a drop while retaining its initial (spherical)
shape will undergo rigid body rotation and align itself along the symmetry
axis \cite{graham_2018}. In a linear shear flow, where these two components are
in equal proportion, the drop would stretch linearly in time and tilt toward the
flow direction \cite{Young2008}.

\begin{figure*}
\subfloat[]{ \centering
\includegraphics[scale=0.4]{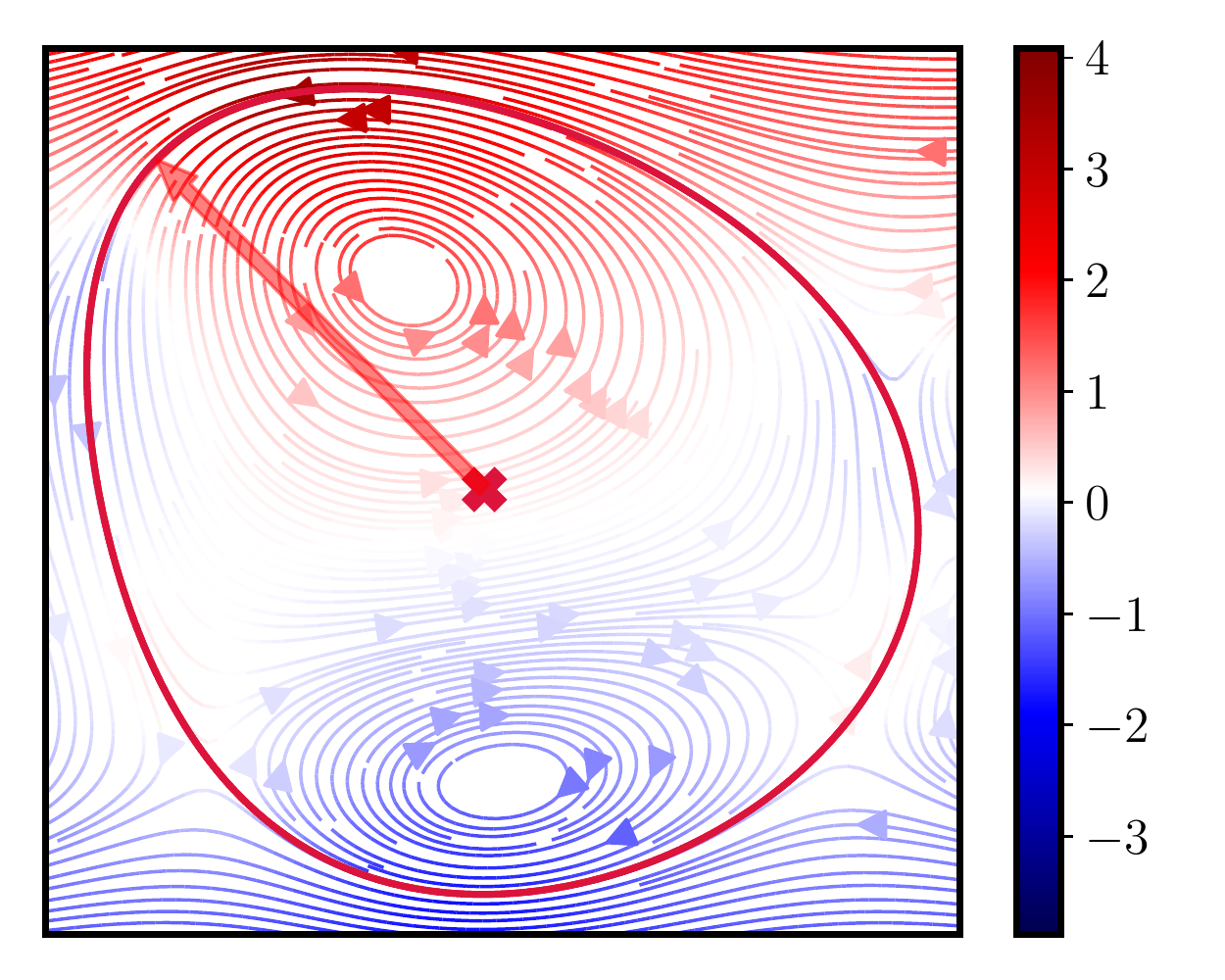}}
\subfloat[]{ \centering
\includegraphics[scale=0.4]{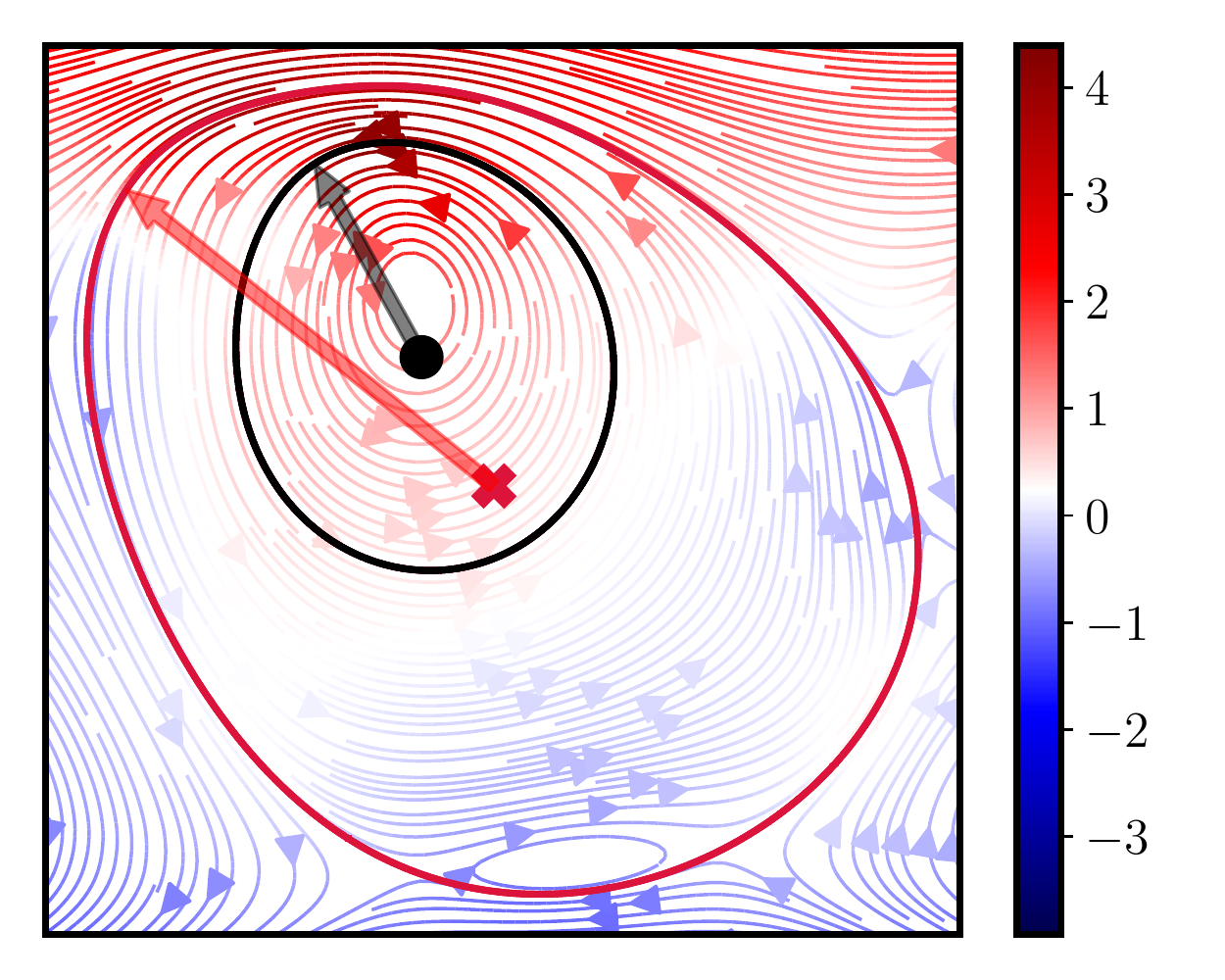}}

\subfloat[]{ \centering
\includegraphics[scale=0.4]{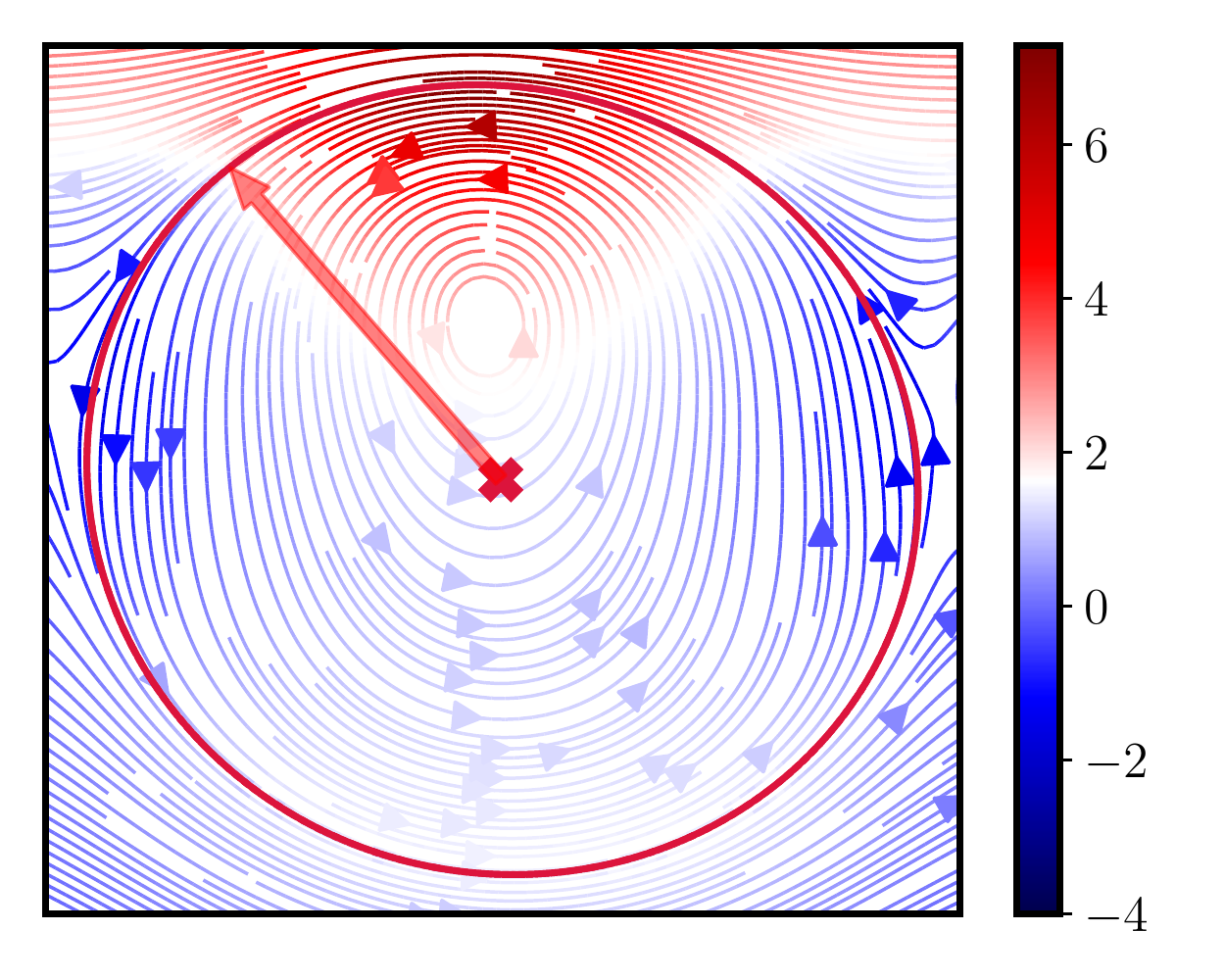}}
\subfloat[]{ \centering
\includegraphics[scale=0.4]{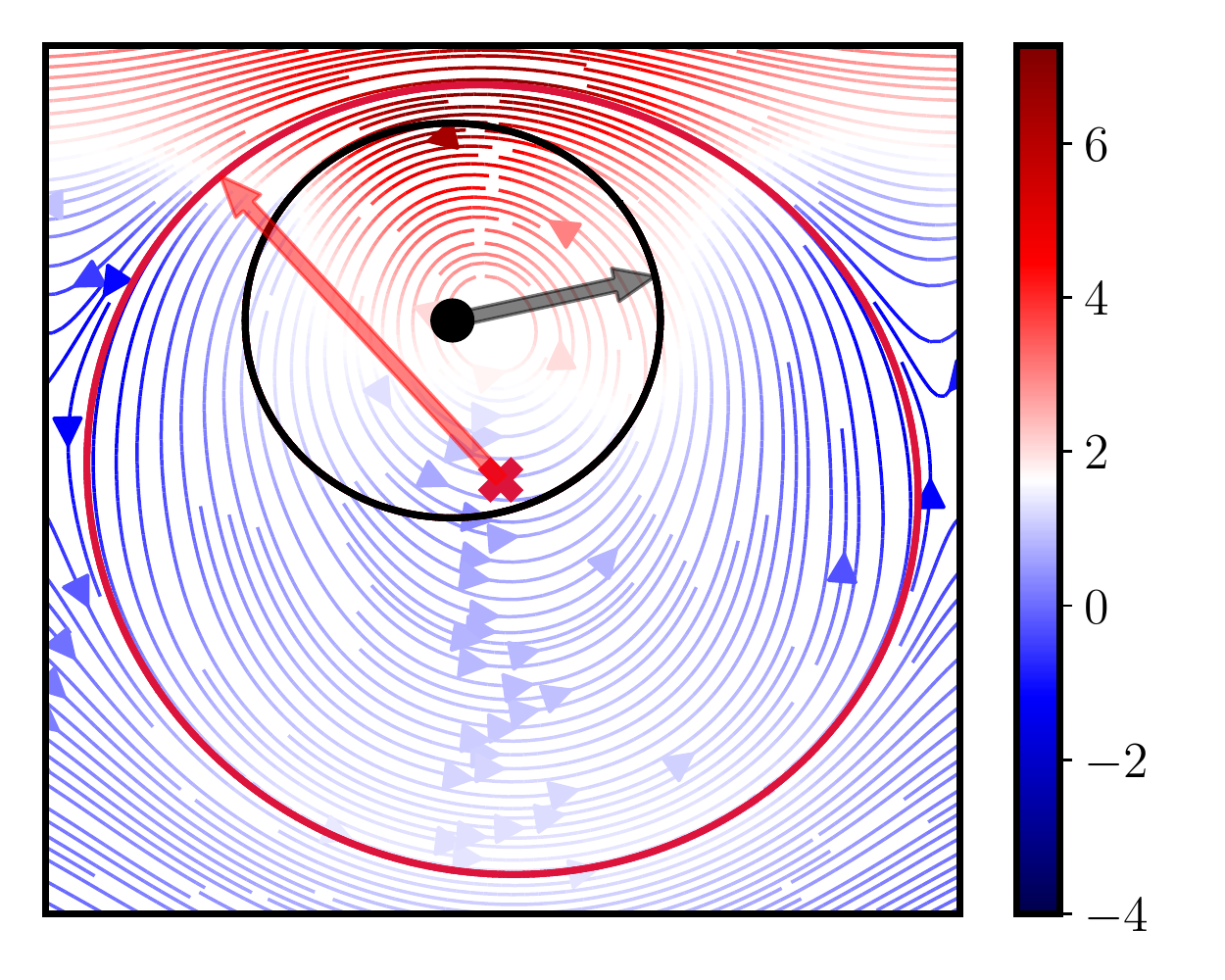}}
\caption{Comparison of streamlines between simple drop and compound drop at
  equilibrium: (a) simple drop (b) compound drop in stable state (c) simple drop
  (d) compound drop in limit-cycle state. The streamlines are colored by the
  vorticity whose magnitude is given by the colorbar. The red cross marker
  represents the centroid of the compound drop, and the red arrow represents the
  extensional axis of the compound drop along which the it experiences maximum
  deformation. Similarly, the black dot and the black arrow represent the
  centroid and the extensional axis of the core, respectively. The parameters
  used are the same as shown in Fig. \ref{fig_3}.}
\label{fig_4}
\end{figure*}

\begin{figure*}
  \subfloat[]{\centering
    \includegraphics[scale=0.35]{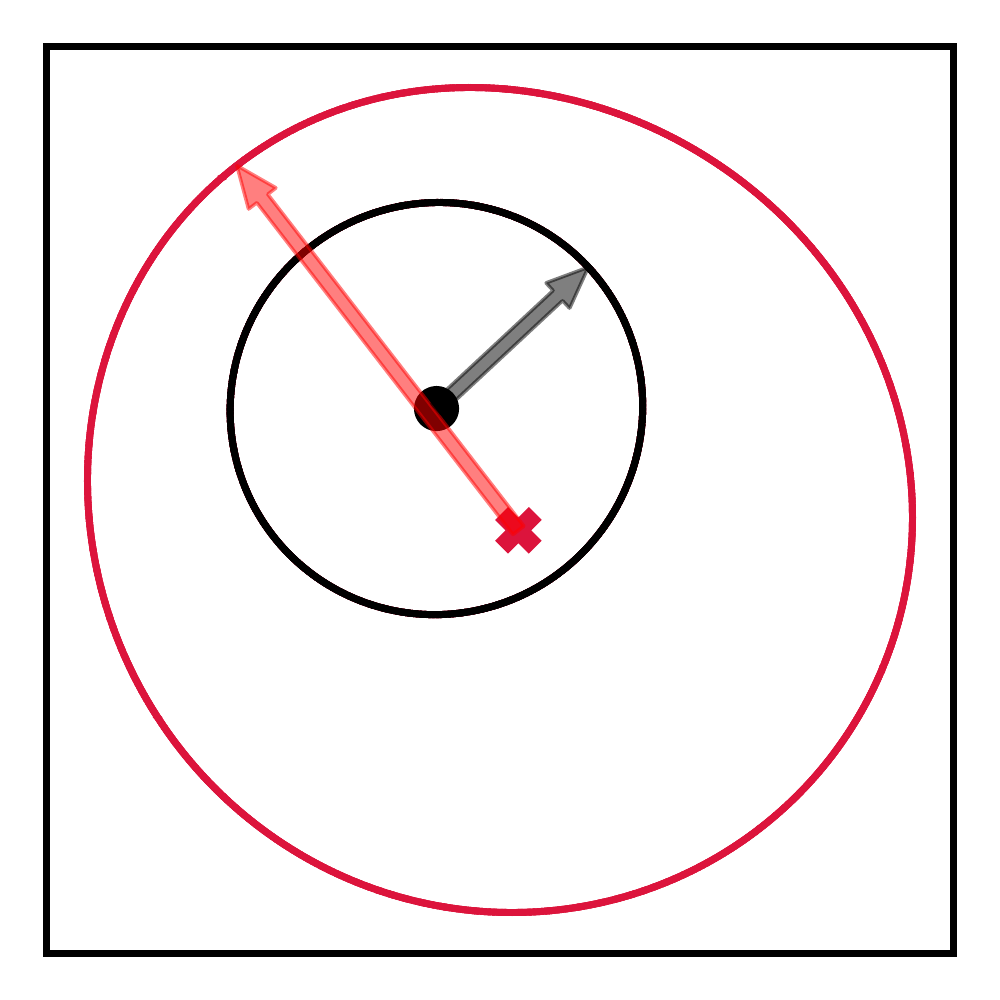}
  }
    \subfloat[]{\centering
    \includegraphics[scale=0.35]{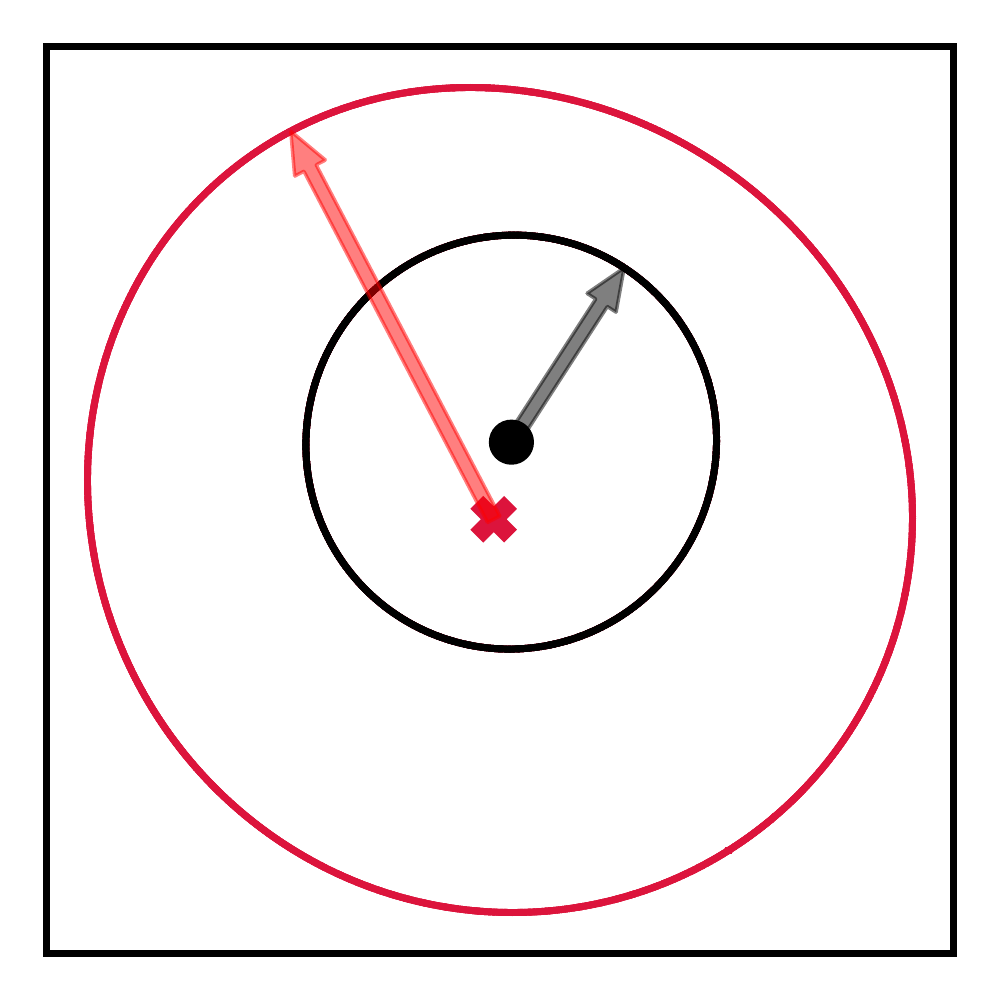}
  }
    \subfloat[]{\centering
    \includegraphics[scale=0.35]{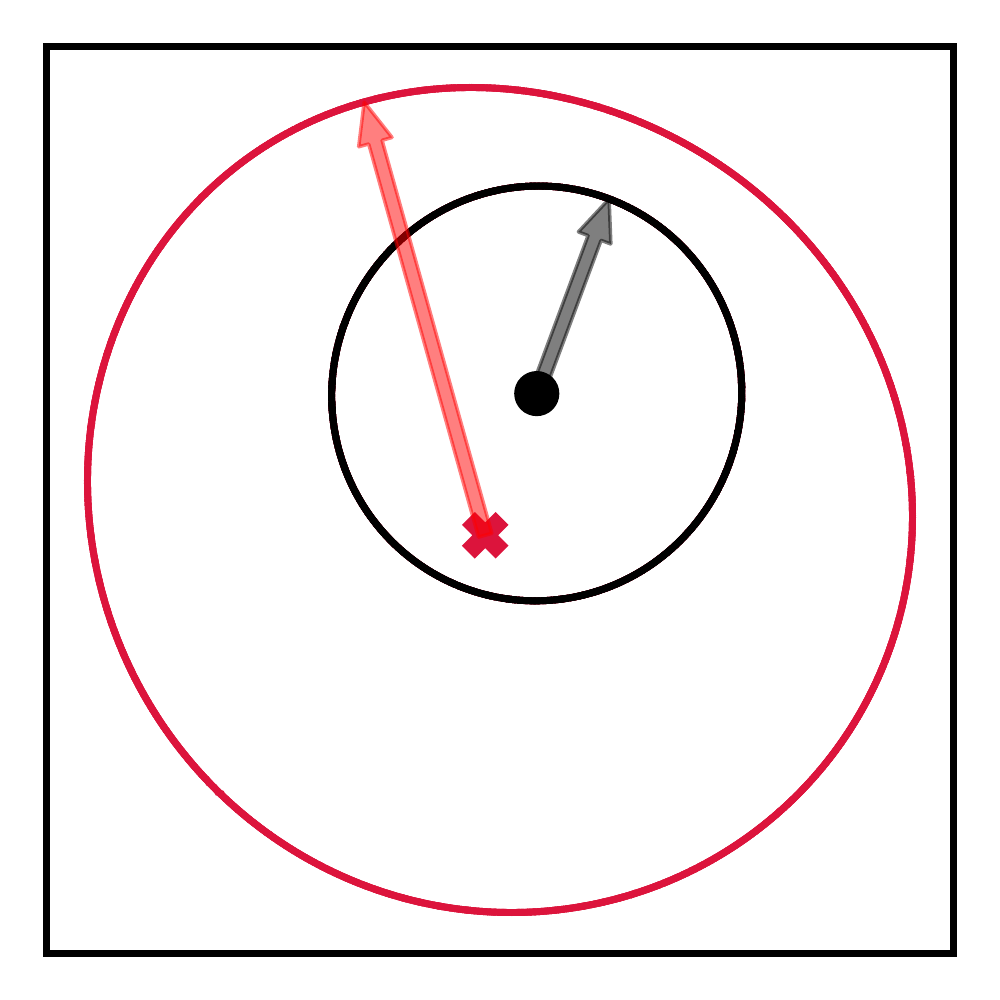}
  }
    \subfloat[]{\centering
    \includegraphics[scale=0.35]{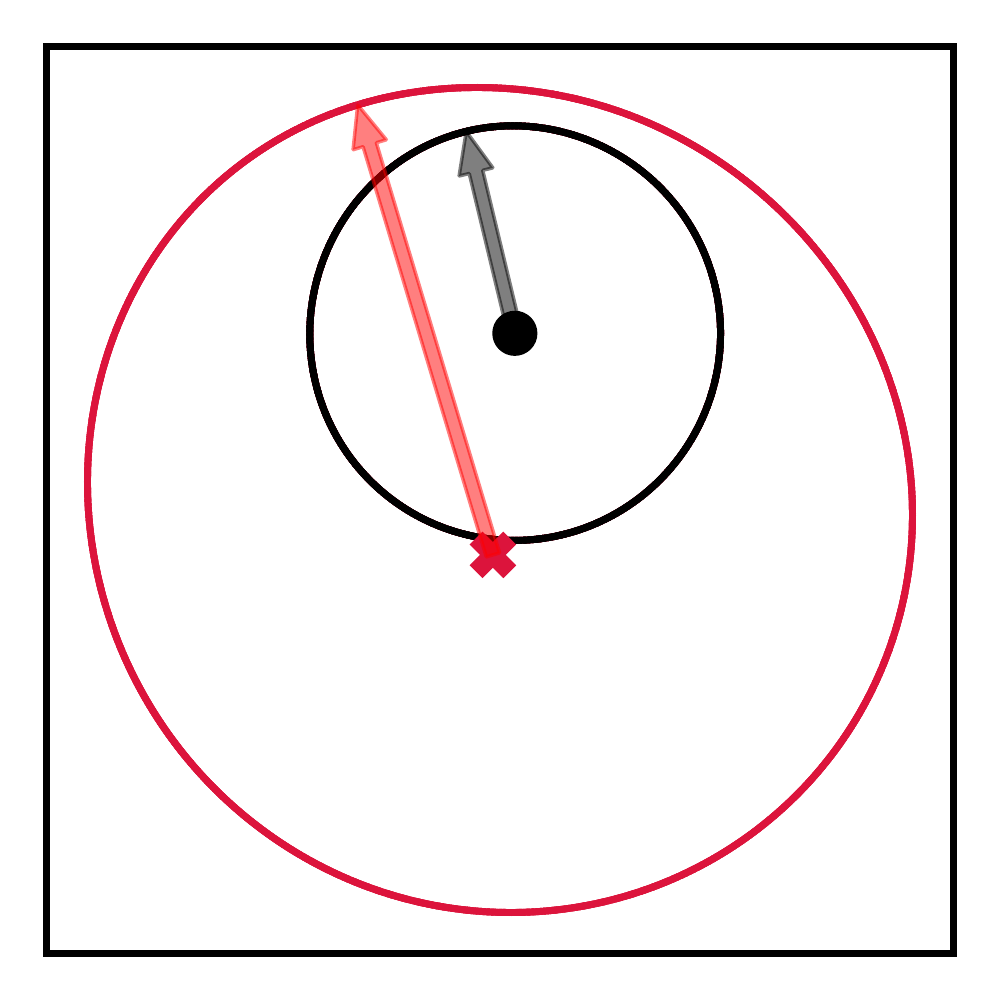}
  }
    \subfloat[]{\centering
    \includegraphics[scale=0.35]{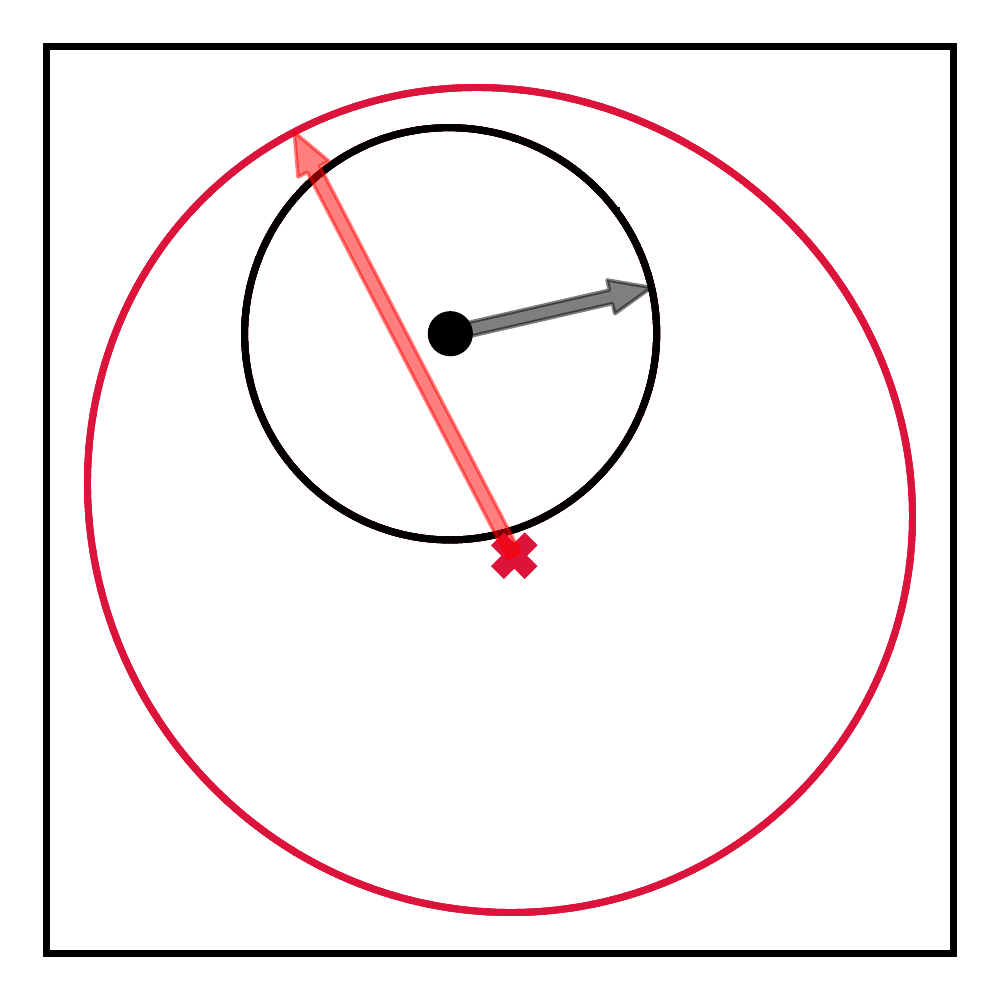}
  }
  \caption{Revolving motion of the core in limit-cycle state. Figures (a)-(e)
represent the position of the core inside the shell at points $A$-$E$ in
Fig. \ref{fig_3}(b). The red cross marker and the black dot represent the
centroids of the compound drop and the core, respectively. The arrows in red and black
colours represent the extensional axis of the compound drop and the core, respectively.}
\label{fig_5}
\end{figure*}

In the Poiseuille flow simulated here, we identify the magnitude of strain rate
tensor from the deformation of the core, and the vorticity tensor from the local
vorticity field. From Fig. \ref{fig_4}(b) and Fig. \ref{fig_4}(d), we see that
the magnitude of the vorticity field in limit-cycle state is higher than in the
stable state. The prolate ellipsoid and the circular shapes of the core in
Fig. \ref{fig_4}(b) and Fig. \ref{fig_4}(d) clearly show that the deformation is
higher in the stable state. This qualitative conclusion is further supported by
the quantitative measurements of the deformation shown in Table \ref{tab_1},
where we observe that the deformation parameter $\mathcal{D}$ for the core in
stable state is one order of magnitude higher than the limit-cycle state. Thus,
large deformation and small vorticity field experienced by the core in stable
state shows that the strain rate tensor is dominant. This dominant strain rate
tensor aligns the core along the principal axes (see Fig. \ref{fig_4}(b))
thereby arresting its revolving motion. The dominant vorticity field in the
limit-cycle state causes the core to revolve in a nearly fixed orbit which
resembles rigid-body rotation. It also causes the extensional axis of the core
along which it is stretched, to oscillate as shown in Fig. \ref{fig_5}.

\subsubsection{Deformation dynamics}
As the core revolves, it comes closer to the shell interface in each cycle. At
this instant, the liquid in the gap between the two interfaces gets squeezed
causing a pressure buildup which affects the deformation of the core every time
it comes near the shell interface.  This pressure buildup decreases in each
cycle as the core moves inward due to increase in the thickness of the
gap. Thus, the deformation of the core characterized by the deformation
parameter $\mathcal{D}$ oscillates as seen in Fig. \ref{fig_6}(a) and
Fig. \ref{fig_6}(b). These oscillations in the case of a stable state decay with
time, and reaches a constant value once the core stops revolving. The
deformation of the core in the limit-cycle state continues to oscillate due to
its revolving motion. The kinks (point D in Fig. \ref{fig_6}(b)) observed in
$\mathcal{D}$ at each cycle are due to the slightly out of phase nature of the
oscillations of the core. As shown in Fig. \ref{fig_5}(d) and Fig
\ref{fig_5}(e), the core reaches its maximum displacement in both directions one
after the other which in turn affect their deformation thereby resulting in the
kinks. Similar trends have also been reported in temporal evolution of the
deformation of the simple drop in Poiseuille flow under Stokes conditions
\cite{Blawzdziewicz2010}.

\subsection{Dynamics of the compound drop}
\subsubsection{Deformation}
\begin{table*}[]
  \begin{tabular}{lllllllllll}
    \hline
    \hline
\multirow{2}{*}{$\lambda$} & \multirow{2}{*}{$K$} & \multirow{2}{*}{$\mu_r$} &
                                                                               \multirow{2}{*}{$Ca$}
  & \multicolumn{3}{c}{Core} & \multicolumn{2}{c}{Compound} & \multicolumn{2}{c}{Simple} \\
                           &                      &                          &
    & \multicolumn{1}{c}{$\epsilon_x$} &  \multicolumn{1}{c}{$\epsilon_y$}   &
                                                                               \multicolumn{1}{c}{$\mathcal{D}$}
                           &  \multicolumn{1}{c}{$y_{d}$}         &  \multicolumn{1}{c}{$\mathcal{D}$}  &  \multicolumn{1}{c}{$y_d$}    &  \multicolumn{1}{c}{$\mathcal{D}$}   \\
    \hline
0.2                        & 0.5                  & 1                        & 0.05                  & 0.035$\pm$0.124 & 0.278$\pm$0.145 & 0.004$\pm$0.003 & 0.672 $\pm$ 0.002 & 0.026$\pm$0.002 & 0.681    & 0.026           \\
0.2                        & 0.5                  & 1                        & 1
    & -0.188       & 0.307          & 0.061         & 0.572           & 0.165 &
                                                                               0.537    & 0.128        \\
    \hline
  \end{tabular}
  \caption{Equilibrium characteristics of the core and the compound drop in
    stable state ($Ca=1$) and limit-cycle state ($Ca=0.05$). The equilibrium
    behavior of the compound drop is also compared with the simple drop under
    the same conditions.  }
\label{tab_1}
\end{table*}

\begin{figure*}
  \subfloat[]{
    \centering
    \includegraphics[scale=0.7]{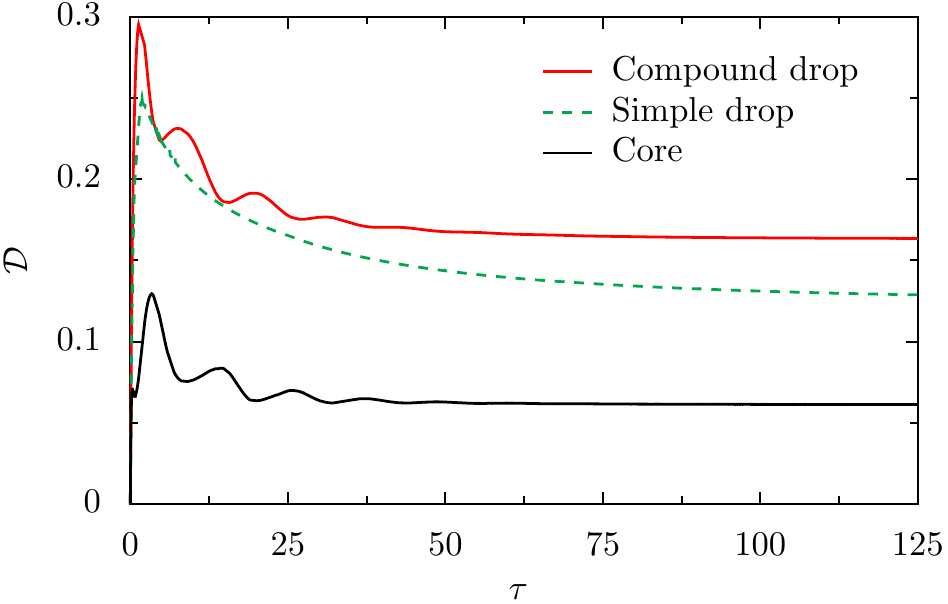}}
  \subfloat[]{
    \centering
    \includegraphics[scale=0.7]{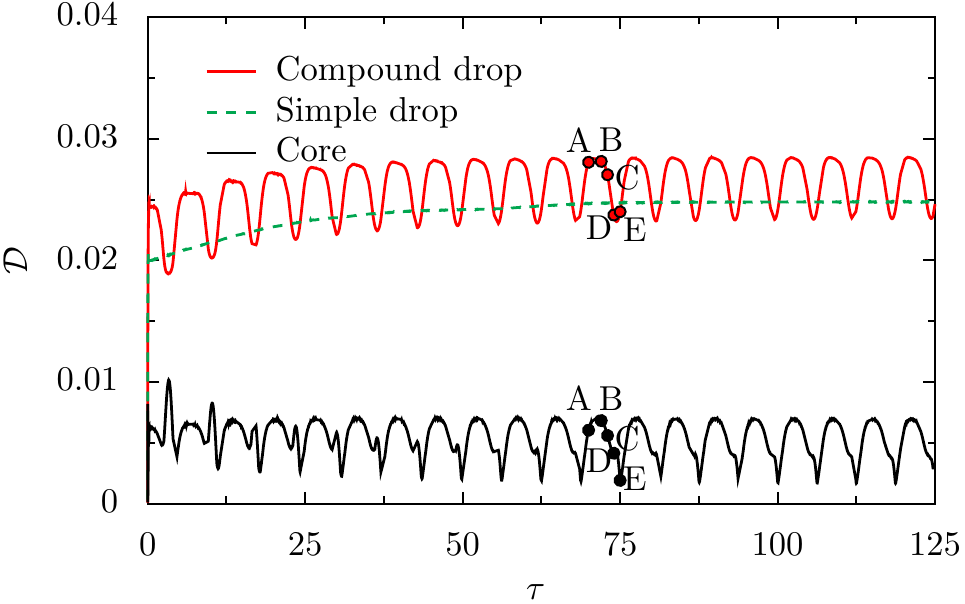}}

\subfloat[]{
    \centering
    \includegraphics[scale=0.7]{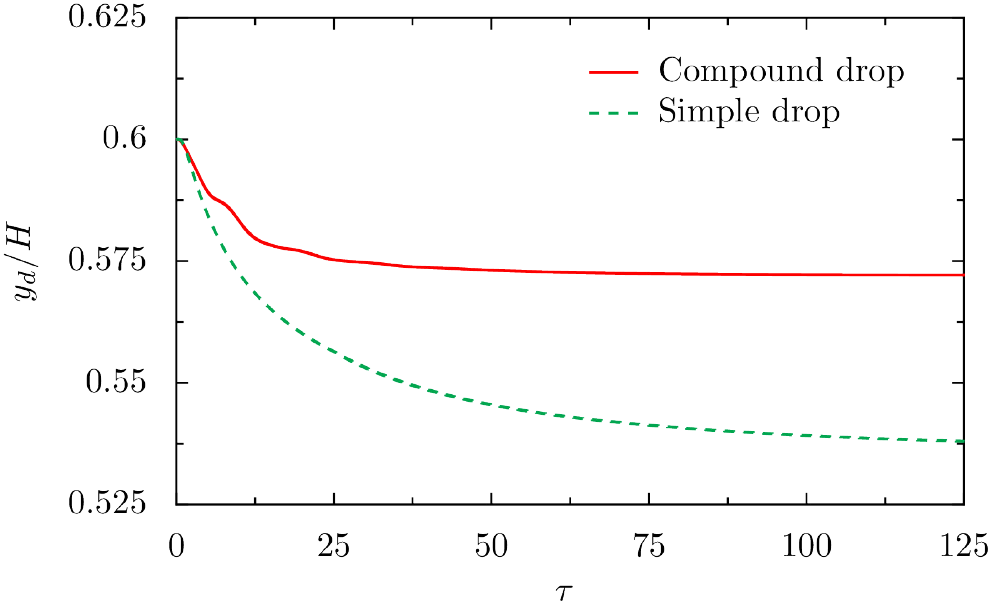}}
  \subfloat[]{
    \centering
    \includegraphics[scale=0.7]{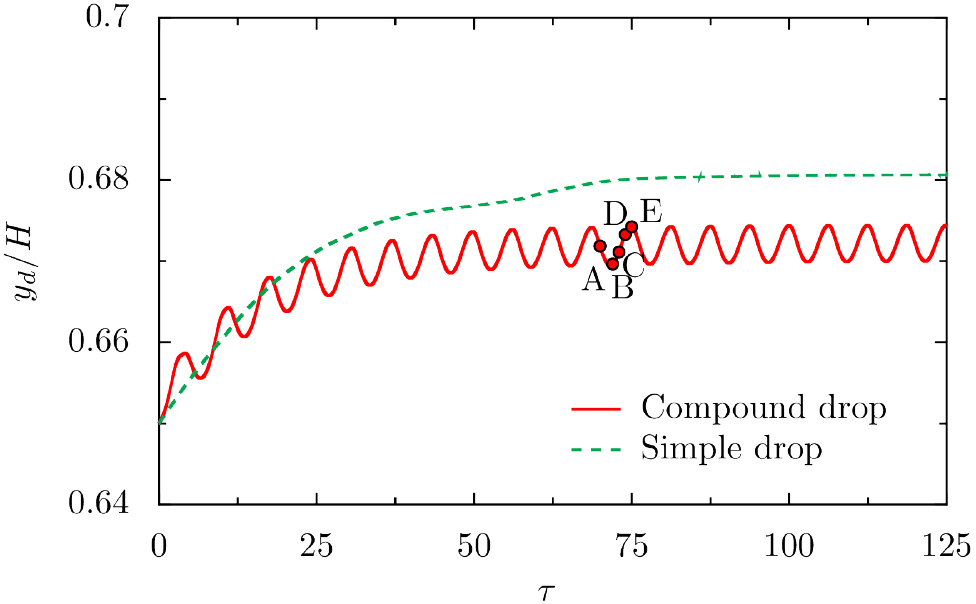}}

  \caption{Temporal evolution of the deformation parameter $\mathcal{D}$ of the
    compound drop and the core under (a) stable state and (b) limit-cycle
    state. Temporal evolution of the trajectory of the compound drop under (c)
    stable state and (d) limit-cycle state. The trajectory and the deformation
    of the compound drop are also compared with the simple drop under the same
    conditions. The parameters used for the two states were
    the same as in the Fig. \ref{fig_3}. }
\label{fig_6}
\end{figure*}

The temporal evolution of the deformation parameter $\mathcal{D}$ of the
compound drop in the two states are shown in Fig. \ref{fig_6}(a) and
Fig. \ref{fig_6}(b). In each state, the oscillations in $\mathcal{D}$ for the
compound drop is very similar to the ones in the core. Similar oscillations in
$\mathcal{D}$ have also been observed for compound drops in linear shear flows
\cite{Chen2015,Chen2015a,Kim2017}. These oscillations are due to the revolving
motion of the core. Thus the parameter $\mathcal{D}$ attains a constant value in
the stable state once the core stops revolving, whereas it continues to
oscillate in the limit-cycle state like the core. As expected, the deformation
of the compound drop in the stable state is significantly larger than in the
limit-cycle state since $Ca$ is larger for the stable state. The deformation of
the compound drop in both states is larger than the core which is expected since
the capillary force is inversely proportional to the drop size.

As explained in the previous section, there is a pressure buildup in the gap
between the core and the shell in each cycle when the core comes closest to the
shell. This pressure buildup results in stretching of the shell which then
relaxes as the core moves away from the shell. This cycle of stretching and
relaxing of the shell resembles breathing, which has also been reported for
compound drops in linear shear flows \cite{Chen2015,Chen2015a}. This breathing
motion is very prominent in the stable state during the initial stages as the
deformation of the compound drop is higher than in the limit-cycle state (see
Fig. 1 and Fig. 2 in the supplementary information).  This cycle of stretching
and relaxing of the compound drop is reflected as oscillations in $\mathcal{D}$
shown in Fig. \ref{fig_6}(a) and Fig. \ref{fig_6}(b). By comparing the core
positions ($A-E$) in Fig. \ref{fig_5} with the corresponding $\mathcal{D}$
values in Fig. \ref{fig_6}(b), we find that the deformation of the compound drop
increases when the core is further away from the shell interface, and decreases
when the core is closer to the interface.

The effect of core on the compound drop is evaluated by comparing the temporal
evolution of $\mathcal{D}$ with the simple drop (see Fig. \ref{fig_6}(a) and
Fig. \ref{fig_6}(b)). The deformation of the compound drop in the stable state
is slightly larger than the simple drop. In the limit-cycle state where
$\mathcal{D}$ continues to oscillate, the maximum $\mathcal{D}$ observed in each
cycle is higher than the simple drop, while its minimum is lower than the simple
drop. In addition to enhancing $\mathcal{D}$, the presence of core also alters
the shape of the compound drop when compared to the simple drop. The parachute
shaped simple drop transforms to an egg shape due to the core in stable state
(see Fig. \ref{fig_4}(a) and Fig. \ref{fig_4}(b)). Both the compound drop in the
limit-cycle state and the simple drop retain their initial circular shape due to
low deformation. The extensional axis of the drop along which the drop is
stretched for both simple and compound drop in stable state are in the same
direction. As show in Fig. \ref{fig_5}, the extensional axis of the compound
drop in limit-cycle state oscillates due to the revolving core. Thus from the
above comparison with the simple drop, we find that the presence of core
enhances the deformation of the compound drop.

\subsubsection{Migration} \label{sec_mig}

In a non-inertial flow, the cross stream migration of a simple drop occurs due
to its deformation. This deformation might be due to either the hydrodynamics of
the flow or the confinement effects from the wall or a combination of both. The
deformation induced lift force always pushes the drop towards the center of the
channel where the deformation is minimum \cite{Mortazavi2000, Stan2011, Lan2012},
and its magnitude is proportional to the deformation. The drop also migrates
towards the wall if $1\leq \mu_r<10$ \cite{Chan1979,Hur2011}. The above results
apply only to small drops ($\lambda < 0.3$); larger drops ($\lambda \geq 0.3$)
always occupy the center due to confinement effects.

We quantify the migration of the compound drop by measuring its height $y_d$
from the bottom wall. The temporal evolution of migration of the compound drop
and the simple drop are shown in Fig. \ref{fig_6}(c) and Fig. \ref{fig_6}(d) for
both the states. The direction of migration of the compound drop is the same as
the simple drop in both states. When compared to a simple drop, the compound
drop oscillates during its migration. These oscillations are markedly noticeable
in the limit-cycle state and continues to do so even after reaching
equilibrium. In the stable state, we observe only minor undulations during its
migration which stops once it reaches its equilibrium position. By comparing the
core positions marked by the points $B,D,E$ in Fig. \ref{fig_3}(b) with the
corresponding location of the drop in Fig. \ref{fig_6}(d), we see that they
reach their maximum and minimum heights together in each cycle. This
synchronized motion is a result of the core pushing the shell upward as it
reaches its peak in each cycle. Similar oscillatory motion due to the core has
been observed experimentally in self propelling encapsulated nematic shells
\cite{Hokmabad2018}.

The equilibrium position of the compound drop is slightly higher than the simple
drop in the stable state whereas it is slightly lower than the simple drop in
the limit-cycle state. The slow upward motion of the core after it stops
revolving in the stable state pushes the compound drop outward resulting in its
slightly higher equilibrium position. The compound drop in the limit-cycle state
is closer to the centerline than the simple drop. This is due to the slightly
larger deformation induced lift force for the compound drop from its maximum
deformation in each cycle, which is larger than the deformation
of the simple drop (see Fig. \ref{fig_6}(b)). The continued
revolving motion of the core in the limit-cycle state also makes the compound
drop to oscillate at its equilibrium position. Thus, the presence of core
affects the migration of the compound drop.

The change in the equilibrium position of the compound drop due to the core also
affects the circulation pattern inside the shell. We know that the circulation
is the result of shearing of the shell interface due to the difference in
velocities between the interface and the surrounding fluid. This shearing
changes with position of the drop since the fluid velocity changes with the
position. From Fig. \ref{fig_4}(a) and Fig. \ref{fig_4}(b), we observe that the
circulation in the bottom half of the simple drop reduces significantly
in the compound drop due to its slightly higher position. Since the compound
drop in the limit-cycle state and its equivalent simple drop are completely in
the upper-half of the channel, we observe from Fig. \ref{fig_4}(c) and Fig.
\ref{fig_4}(d) only a slight shift in the center of the circulation. At
equilibrium, when the core has stopped revolving in stable state, it migrates to
the center of the circulation region.

\begin{figure}
\centering
      \includegraphics[scale=2]{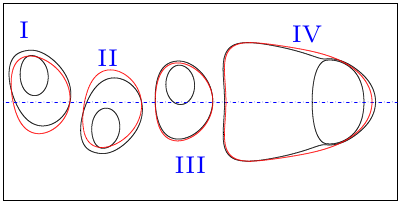}
      \caption{Effect of migration of the compound drop on the eccentricity of
        the core. The interface shapes shown in red and black colours represent
        the simple and compound drops, respectively. The blue dash-dotted line
        represents the centerline of the channel. The parameters used for these
        cases are given in the Table \ref{tab_2}.}
   \label{fig_7}
 \end{figure}
 
 \begin{table}[h]
   \begin{threeparttable}
  \begin{tabular}{llllcll}  
    \hline
    \hline
Index & $\lambda$ & $K$ &$\mu_r$ & $Ca$ &\multicolumn{2}{c}{$y_{eq}/H$} \\
\cmidrule(r){6-7}
 && & && Simple    & Compound \\
\hline
I &  0.2 & 0.5 & 1 & 1 & 0.537    & 0.572\tnote{1}      \\
II &  0.2 & 0.5 & 1 & 1 & 0.464    & 0.427\tnote{1}      \\
III &  0.2 & 0.5 & 0.1 & 1 & 0.5    & 0.509      \\
IV &  0.4 & 0.5& 1 & 1 & 0.5    & 0.5      \\
    \hline
    \hline
  \end{tabular}
  \begin{tablenotes}
\item [1] Initial release heights for drops I and II are 0.6 and 0.4, respectively.  \end{tablenotes}
\end{threeparttable}
\caption{Parameters used for cases shown in Fig. \ref{fig_7}.}
\label{tab_2}
\end{table}

By altering the circulation inside the shell, the migration of the compound drop
also influences the position of the core i.e., eccentricity. When there are
multiple circulation regions inside the shell, the core is always driven by the
circulation dominant in size and magnitude. Thus, when the compound drop is
off-centered, the core occupies an asymmetric position. Different core
configurations due to the migration of the compound drop are shown in
Fig. \ref{fig_7}. The parameters used for the different configurations and the
corresponding equilibrium positions are listed in the Table \ref{tab_2}. The
eccentric positions of the core shown in Fig. \ref{fig_7} are often observed in
experiments \cite{Okushima2004,Chen2011,Kemna2012,Hati2016}. It is well known
that for a drop moving in a channel (2D) two equilibrium locations exists on
either side of the centerline. From drops $I$ and $II$ in Fig. \ref{fig_7} we
find that the position of the core switches to opposite sides depending upon the
equilibrium position of the compound drop (either in top or bottom half of the
channel). Despite the different off-centered configurations, the core always
stays closer to its nearest wall. The core becomes symmetric about the
centerline only when the compound drop is also at the center similar to the
drops under axisymmetric conditions
\cite{Zhou2008,Song2010,Tao2013,Borthakur2018,Che2018}. It is well known that
when the drop is at the centerline, there will be two circulation regions
rotating in opposite directions and occupying each half of the drop. These
oppositely moving circulation pushes the core towards the apex of the shell as
shown by the drop $IV$ in Fig. \ref{fig_7}. A
slight deviation of the drop's position from the centerline alters the flow
inside the shell making one of the circulations dominant thereby resulting in an
asymmetric core position as shown by the drop $III$ in Fig. \ref{fig_7}.

The concentric drop configuration used as initial condition in our simulations
is rarely realized in experiments. Hence, we studied the effect of initial
eccentricity of the core on the migration and deformation dynamics of the core
and the compound drop (see supplementary information). In addition, we also
studied the effect of the initial release height of the drop. Based on the
results, we find that the initial conditions only affect their transient
dynamics, while their equilibrium behavior remains the same.

\subsection{Core-shell interaction}

From the results discussed above, we find that the core and the shell interact
through the liquid between them as follows. The ambient flow conditions inside
the channel deforms the compound drop by shearing its interface, and the
resulting deformation induced lift force causes the compound drop to
migrate. The shearing of the shell interface causes the liquid inside it to
circulate, which makes the core to revolve. This revolving core, when it comes
closer to the shell in each cycle, squeezes the liquid in the gap between them
thereby creating a pressure buildup. Through this pressure buildup in each
cycle, the core causes the compound drop to oscillate as it migrates, and also
to undergo breathing motion. As the compound drop migrates towards its
equilibrium position, the continuous change in its position also alters the
circulation inside it due to the change in shearing rate of the shell. The
changing circulation affects the revolving motion of the core which in turn
affects the compound drop dynamics. Once the drop reaches its equilibrium
position, the shearing force on the shell becomes steady and thus a steady
circulation inside it. The nature of the rotational flow field inside the shell
determines the equilibrium behavior of the core. If the strain rate tensor is
dominant, we observe stable state where the core stops revolving and starts
translating outwards very slowly. Limit-cycle state behavior occurs if the
vorticity tensor is dominant, where the core continues to revolve in a fixed
orbit. Thus, the presence of core in both states enhance the deformation of the
compound drop and also affect its migration. The compound drop in turn affects
the position of the core inside the shell and also its equilibrium behavior by
altering the circulation. Thus, we find that the core-shell interaction affects
both the core and the compound drop dynamics.

\subsection{Parametric Study}
In this section, we investigate the influence of each parameter on the
core-shell interaction and their effects on the core and the compound drop
dynamics. Since the transient dynamics depend strongly on the initial
conditions, we only report the equilibrium behavior. We represent the
oscillating behavior of the limit-cycle state cases by taking its average with
error bars which represents the maximum and minimum values.

\subsubsection{Effect of viscosity ratio and capillary number}
\begin{figure*}
  \subfloat[]{
    \centering
    \includegraphics[scale=0.5]{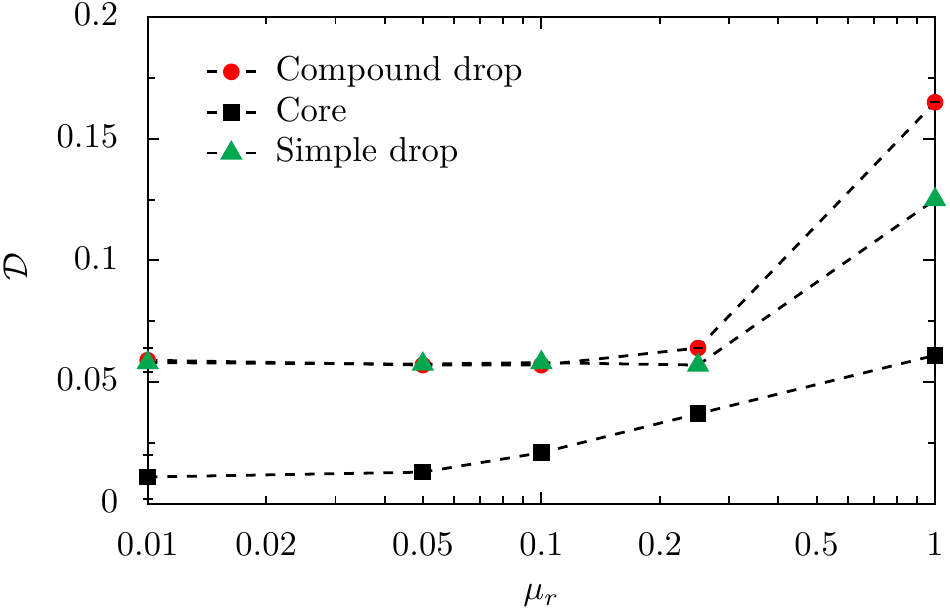}
  }
  \subfloat[]{
    \centering
    \includegraphics[scale=0.5]{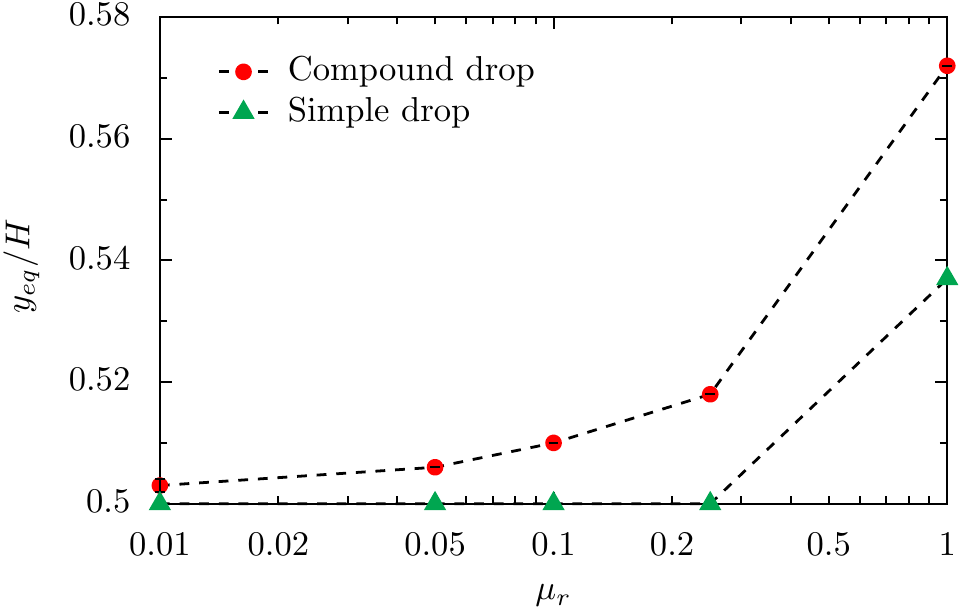}
  }
 \subfloat[]{
    \centering
    \includegraphics[scale=0.5]{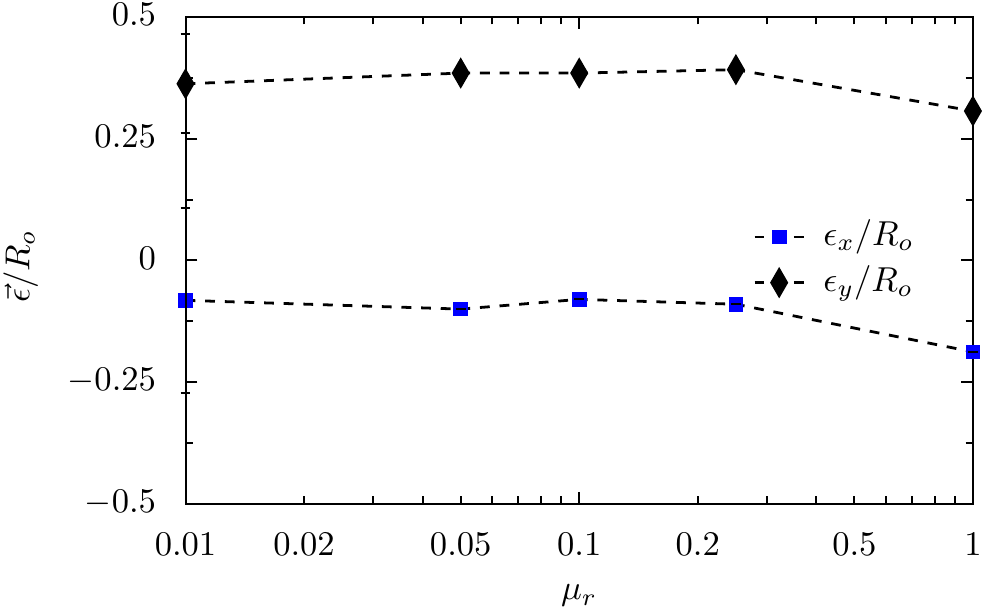}
  }

\caption{Effect of viscosity ratio on the (a) deformation of the compound drop
and the core, (b) migration of the compound drop, and (c) eccentricity of the
core at equilibrium. Other parameters were kept fixed at $\lambda=0.2$, $K=0.5$,
and $Ca=1$.}

  \label{fig_VR}
\end{figure*}

\begin{figure*}
\subfloat[]{
    \centering
    \includegraphics[scale=0.5]{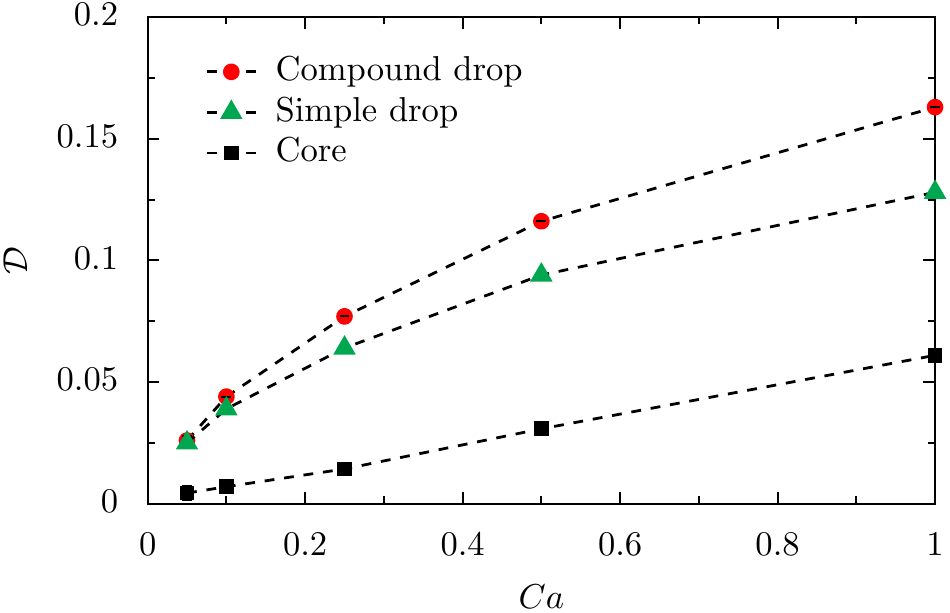}
  }
  \subfloat[]{
    \centering
    \includegraphics[scale=0.5]{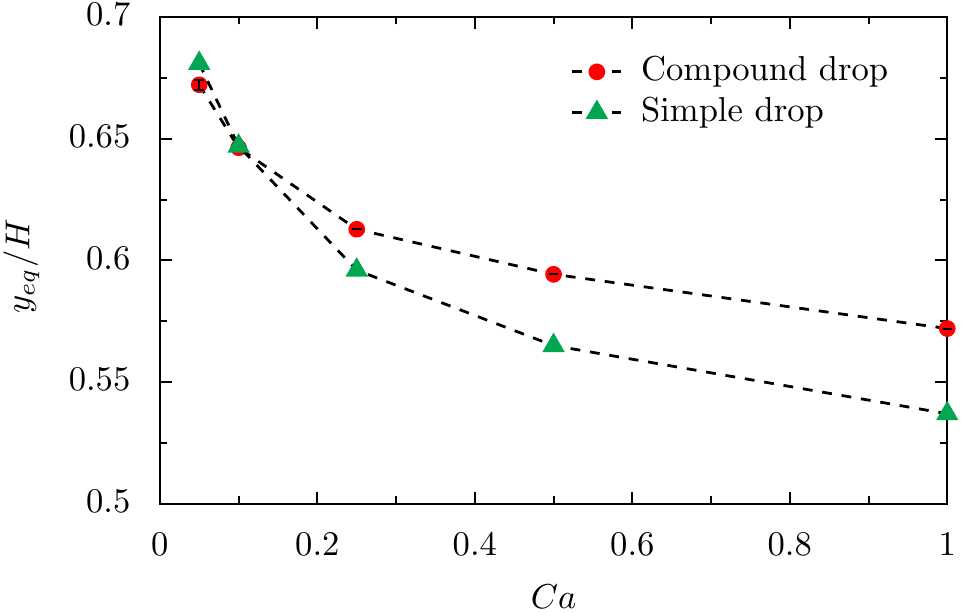}
    }
 \subfloat[]{
    \centering
    \includegraphics[scale=0.5]{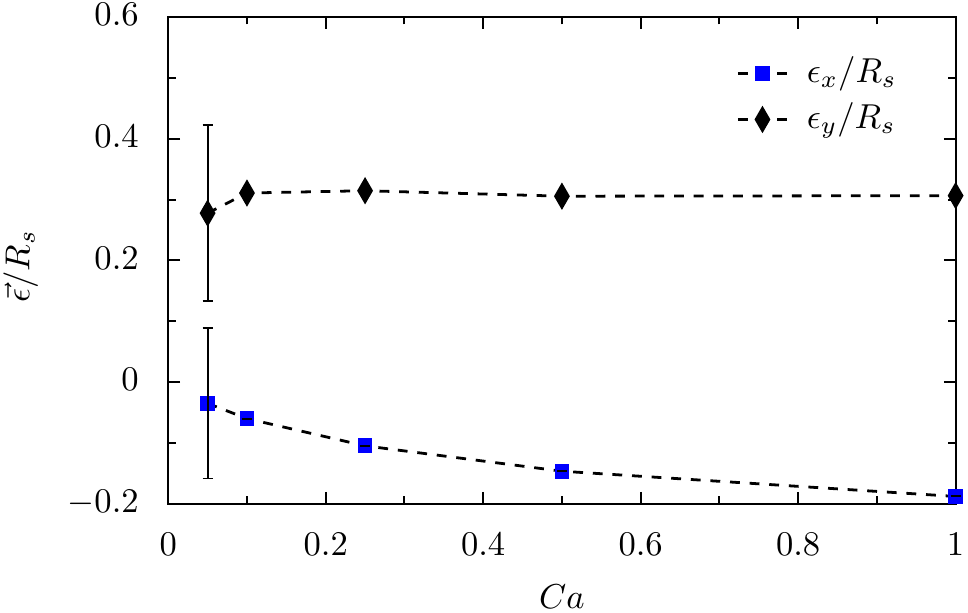}
  }
  \caption{Effect of capillary number on the (a) deformation of the compound
drop and the core, (b) migration of the compound drop, and (c) eccentricity of
the core at equilibrium. Other parameters were kept fixed at $\lambda=0.2$,
$K=0.5$, and $\mu_r=1$.}
  \label{fig_Ca}
\end{figure*}

The effect of viscosity ratio and capillary number on the equilibrium behavior
of the compound drop and the core are shown in Fig. \ref{fig_VR} and
Fig. \ref{fig_Ca}, respectively. The parameter values $\lambda=0.2$, $K=0.5$,
and $Ca=1$ were used while varying viscosity ratio, and $\lambda=0.2$, $K=0.5$,
and $\mu_r=1$ were used while studying the effect of $Ca$.

The deformation of the compound drop characterized by the parameter
$\mathcal{D}$ are shown in Fig. \ref{fig_VR}(a) and Fig. \ref{fig_Ca}(a) for
varying viscosity ratio and capillary number, respectively. We observe that the
deformation of the compound drop remains nearly the same as the simple drop till
$\mu_r=0.1$, and starts increasing thereafter till $\mu_r=1$. The nearly
constant deformation of the drops at low viscosities have also been
experimentally observed for drops moving circular tubes\cite{Olbricht1992}. As
expected, similar to the simple drop, the deformation of the compound drop also
increases with increasing $Ca$. The presence of core enhances the deformation of
the compound drop, and thus it is always higher than the simple drop except at
low values of $Ca$ and $\mu_r$, where it is nearly the same as the simple
drop. The deformation of the core increases with the increase in either $\mu_r$
or $Ca$. Limit-cycle state behavior is observed only for low values in $\mu_r$
and $Ca$. Thus, we observe a transition from limit-cycle state to stable state
with increase in either $\mu_r$ or $Ca$.

The variation of the equilibrium position of the compound drop with $\mu_r$ and
$Ca$ are shown in Fig. \ref{fig_VR}(b) and Fig. \ref{fig_Ca}(b). As mentioned
before, similar to the simple drop, we also observe the compound drop to move
towards the center if $\mu_r< 1$ and to the wall if $\mu_r>1$. The compound
drop moves towards the center with increase in $Ca$ due to the increase in
deformation. The equilibrium position of the compound drop is always higher than
the simple drop in the range of $\mu_r$ studied. Similar observations can be
made with $Ca$ but only at higher values where we observe stable state
behavior. At low $Ca$, the equilibrium position of the simple drop is slightly
higher than the compound drop due to its limit-cycle behavior as explained in
section \ref{sec_mig}.

The variation of the eccentricity of the core as shown in Fig. \ref{fig_VR}(c),
where the eccentricity in both directions increases a little till
$\mu_r=0.25$, and starts decreasing at a relatively higher rate thereafter. The
eccentricity of the core (see Fig. \ref{fig_Ca}(c)) in the vertical direction
remains nearly constant with increase in $Ca$, while in the horizontal direction it
decreases with increase in $Ca$. These trends in the eccentricity can be
understood using the migration of the compound drop. When the drops are
off-centered, the core occupies the backward part of the shell either at the top
(see drop $I$ in Fig.\ref{fig_7}) or at the bottom (see drop $II$ in
Fig. \ref{fig_7}) of the shell depending upon the equilibrium position of the
drop. As the centroid of the compound drop moves closer to the center, the core
starts moving forward to the apex of the shell with its height $\epsilon_y$ (see
drop $III$ in Fig. \ref{fig_7}) gradually decreasing till it reaches the
centerline.

\subsubsection{Effect of aspect ratio}
\begin{figure*}
      \subfloat[]{
    \centering
    \includegraphics[scale=0.5]{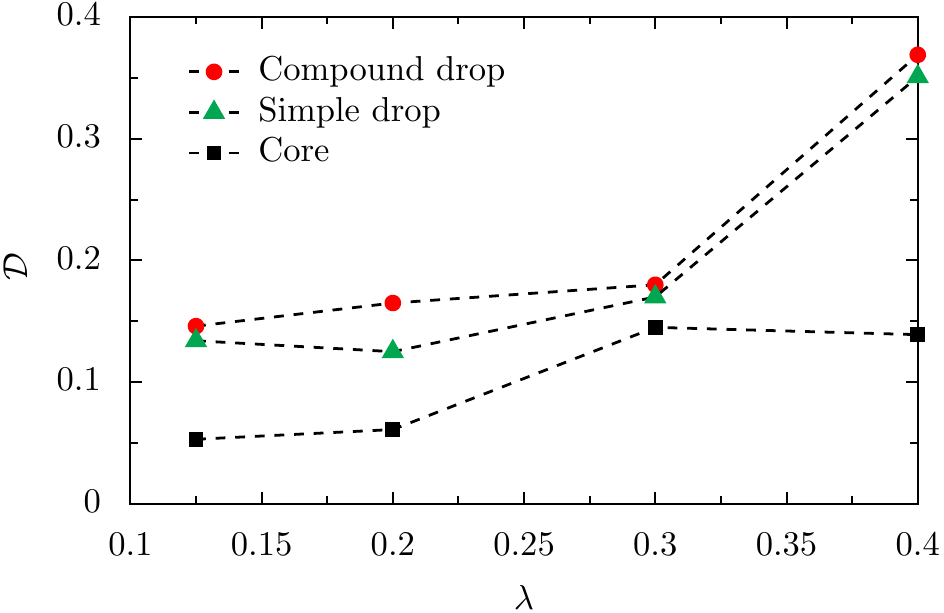}
  }
  \subfloat[]{
    \centering
    \includegraphics[scale=0.5]{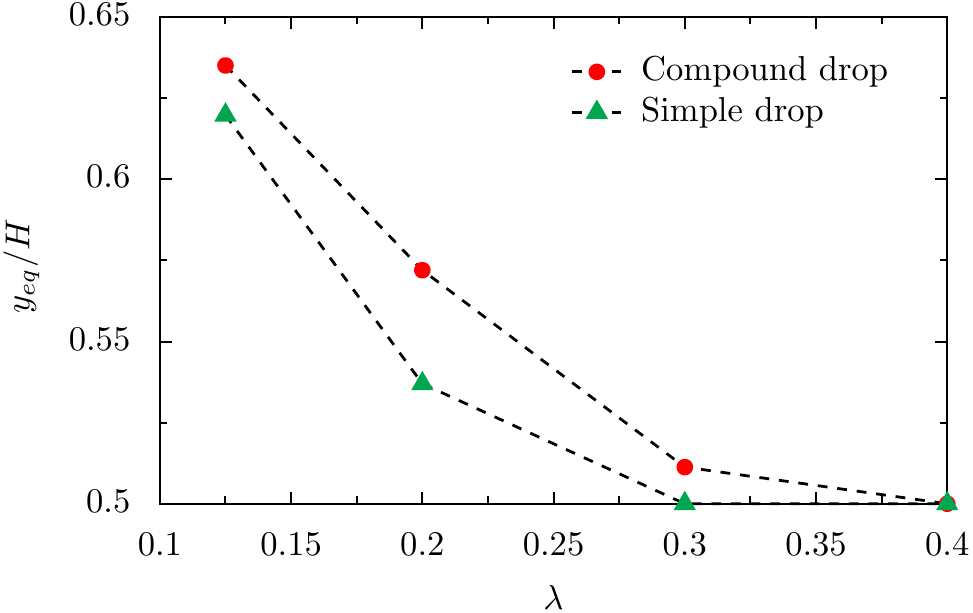}
    }
  \subfloat[]{
    \centering
    \includegraphics[scale=0.5]{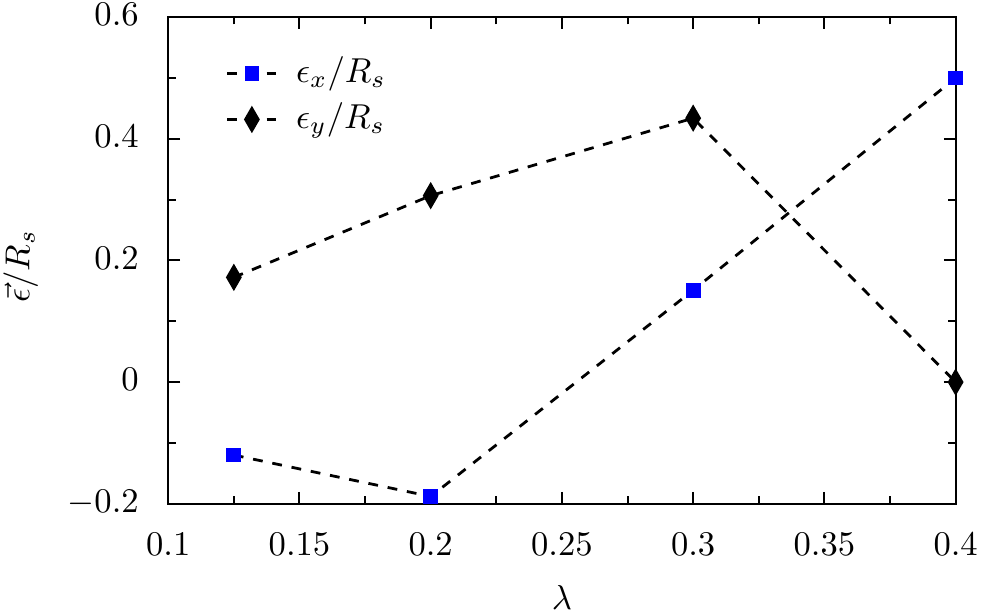}
  }

    \caption{Effect of aspect ratio on the (a) deformation of the compound drop
and the core, (b) migration of the compound drop, and (c) eccentricity of the
core at equilibrium. Other parameters were kept fixed at $K=0.5$, $\mu_r=1$ and
$Ca=1$.}
\label{fig_apr}
\end{figure*}

Figure \ref{fig_apr} shows the effect of aspect ratio on the equilibrium
behavior of the compound drop and the core with $K=0.5$, $Ca=1$, and
$\mu_r=1$. The deformation of the compound drop, as shown in
Fig. \ref{fig_apr}(a), increases with the drop size due to the additional
deformation from the confinement effects. The deformation of the compound drop
is higher than the simple drop due to the core. The deformation of the core also
increases with $\lambda$ but decreases slightly once the compound drop is at the
center, which happens when $\lambda > 0.3$. This reduction in deformation of the
core is probably due to its position at the apex of the shell (see drop IV in
Fig. \ref{fig_7}), and has also been reported for compound drops moving in a
circular pipe \cite{Che2018}. We observed only stable state behavior in the
entire range studied since $Ca$ was fixed at 1, which makes the rate of strain
tensor component dominant inside the shell. At lower $Ca$, we expect limit-cycle
state in smaller drops since they deforms less compared to large drops.

The variation of equilibrium position of the compound drop with its size is
shown in Fig. \ref{fig_apr}(b). The compound drop moves towards the center with
increase in size similar to the simple drop. While the simple drop reaches
center when $\lambda \geq 0.3$, the compound drop reaches the center only when
$\lambda >0.3$. This slight increase in threshold size for reaching the center
is due to the presence of the core, which pushes the compound drop towards the
wall.

The variation of eccentricity of the core with $\lambda$ is shown in
Fig. \ref{fig_apr}(c). The eccentricity in the vertical direction initially
increases with $\lambda$, but reaches zero when the compound drop is in the
center. The eccentricity in horizontal direction initially decreases and starts
increasing reaching its maximum when the compound drop is at the center. The
core as shown in Fig. \ref{fig_7} occupies backward part of the shell at the
top, and as the drop moves towards the center the core starts moving forward to
the front slowly, which explains the increasing trend in eccentricity.

\subsubsection{Effect of radius ratio}

\begin{figure*}
  \subfloat[]{
    \centering
    \includegraphics[scale=0.5]{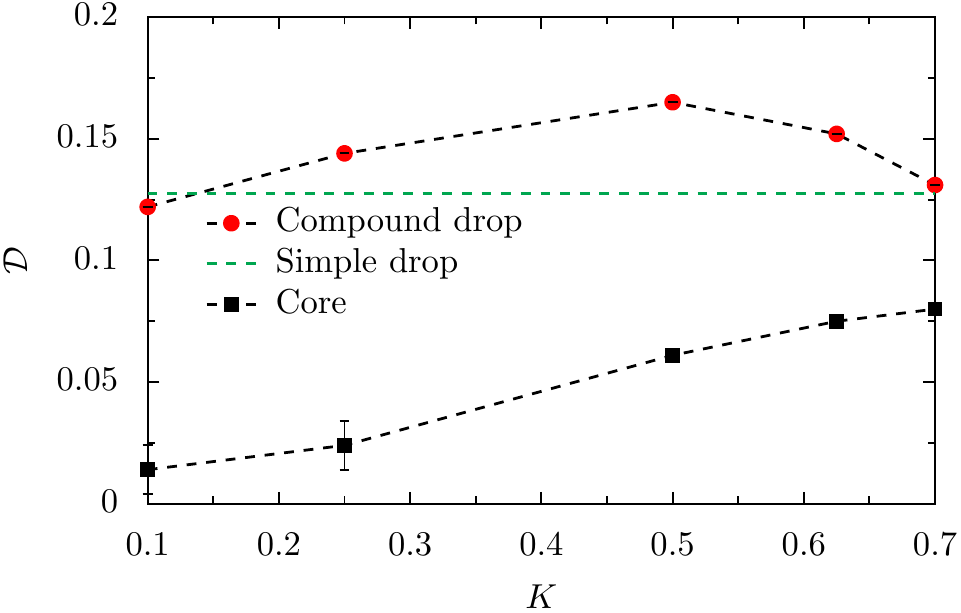}
  }  
  \subfloat[]{
    \centering
    \includegraphics[scale=0.5]{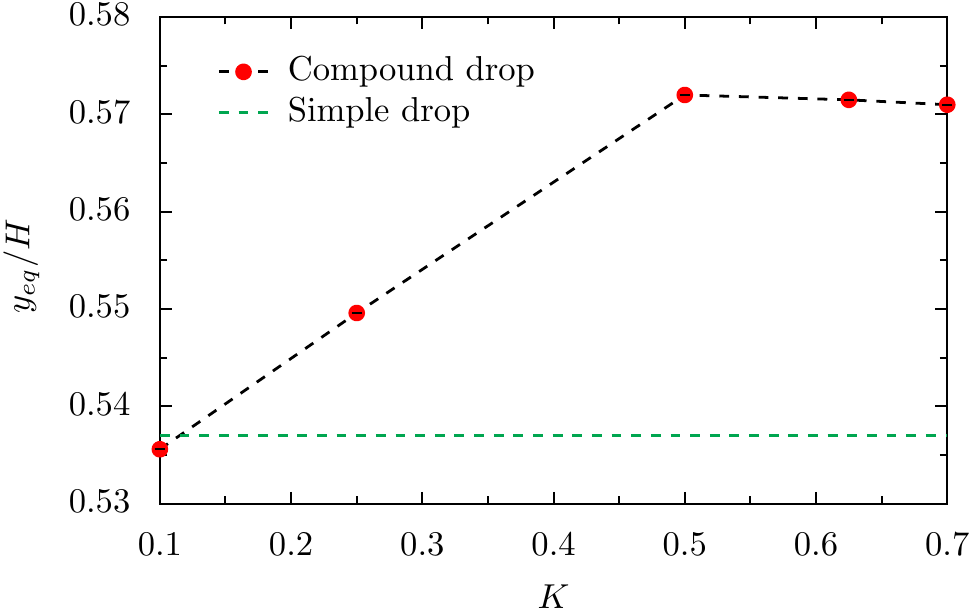}
    }
\subfloat[]{
    \centering
    \includegraphics[scale=0.5]{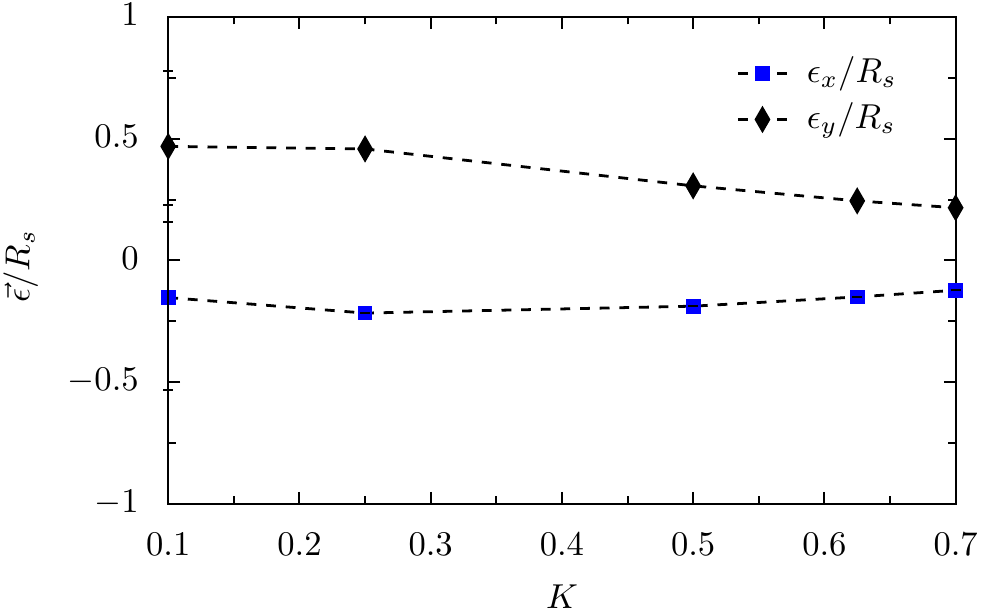}
  }
  \caption{Effect of radius ratio on the (a) deformation of the compound drop
and the core, (b) migration of the compound drop, and (c) eccentricity of the
core. Other parameters were kept fixed at $\lambda=0.2$, $\mu_r=1$ and $Ca=1$.}
  \label{fig_RR}
\end{figure*}

Figure \ref{fig_RR} shows the effect of radius ratio on the equilibrium behavior
of the compound drop and core with $\lambda=0.2$, $Ca=1$, and $\mu_r=1$. The
deformation of the compound drop as shown in Fig. \ref{fig_RR}(a), increases
with core size till $K=0.5$ and starts decreasing till $K=0.7$, where
it is nearly the same as the simple drop. At the smallest core size ($K=0.1$)
simulated, we also find that the deformation of the compound drop slightly
smaller than the simple drop. Such reduction in $\mathcal{D}$ at extreme core
sizes have also been observed for compound drops moving in circular pipes
\cite{Che2018}, and in shear flows
\cite{Hosseini2012,Hua2014,Chen2015,Chen2015a,Luo2015}. The reduction in
deformation at small core sizes occurs due to the low interaction between the
core and the shell \cite{Hua2014}. Larger cores tend to force the shell interface
conform to its shape which affects the local strain inside the shell, and this
reduces the deformation of the compound drop\cite{Hosseini2012}. The
deformation of the core, however, increases monotonically with its size due to
the decrease in the capillary force. Thus, we observe a transition from
limit-cycle to stable state behavior with increase in core size.

The equilibrium position of the compound drop increases with the core size as
seen in Fig. \ref{fig_RR}(b), which indicates that the drop moves
increasingly towards the wall. This behavior with core size confirms that the
presence of core pushes the compound drop towards the wall. The equilibrium
position is slightly lower than the simple drop when the core size is smallest
due to its limit-cycle behavior.

The eccentricity of the core as shown in Fig. \ref{fig_RR}(c), decreases in
vertical direction and increases in horizontal direction with $K$. This
decreasing trend shows that the core moves towards the center. As the core
becomes larger it forces the shell to conform to its shape, which makes the
centroid of the core gradually move closer to the centroid of the compound drop.

\begin{figure}[h]
  \centering
  \includegraphics[scale=0.7]{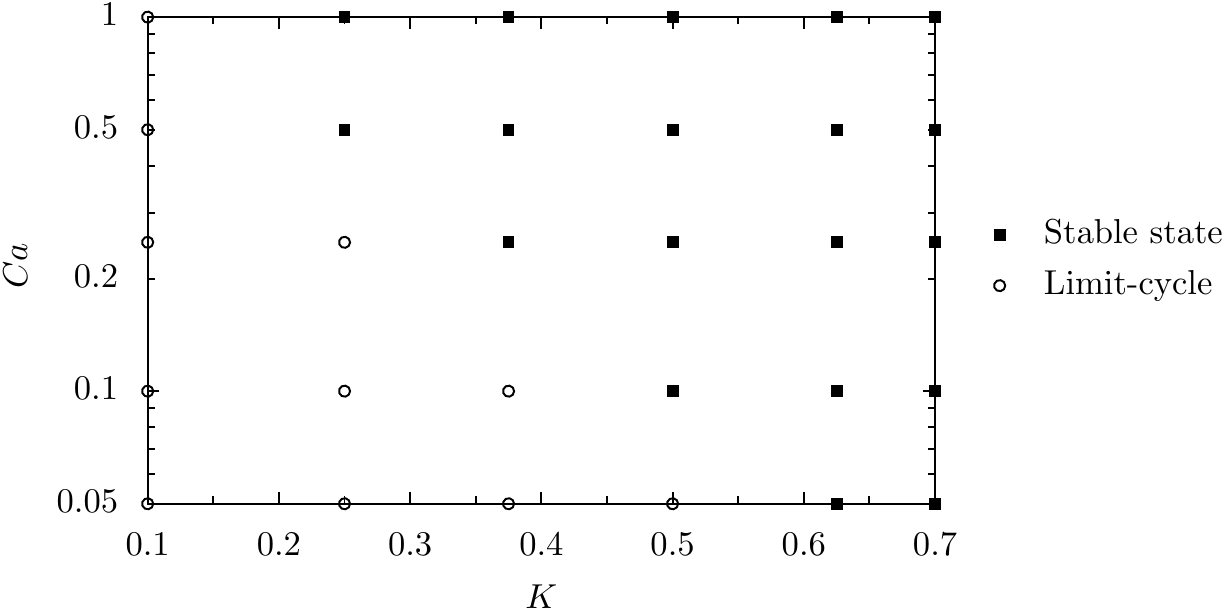}
  \caption{Regime map showing the two equilibrium states in $(K,Ca)$ space. Other parameters were kept fixed at $\lambda=0.2$ and $\mu_r=1$.}
  \label{fig_map}
\end{figure}

Based on the simulations, we have constructed a regime map of $K$ versus $Ca$
with $\lambda = 0.2$ and $\mu_r =1$, which shows the two equilibrium core
behaviors and their transition from one to the other. From Fig. \ref{fig_map},
we observe that the transition from limit-cycle state to stable state behavior
occurs with increasing either $K$ or $Ca$. We also find that the threshold $Ca$
above which the stable state is observed increases with decrease in $K$. This
behavior is expected since smaller cores undergo less deformation compared to
larger cores at a given capillary number.

\section{Summary}

We have investigated numerically the hydrodynamics of a compound drop suspended
in a Poiseuille flow under Stokes regime. Our results show that the core and
shell interact and influence each other through the liquid in the gap between
them. The core starts revolving about the centroid of the compound drop
initially due to the circulation inside the shell. At equilibrium, we identify
two distinct equilibrium core behaviors: stable state and limit-cycle state. In
the stable state the core stops revolving and starts moving outward very
slowly. The core in the limit-cycle state, continues to revolve at a constant
rate and in nearly a fixed orbit. By comparing with the simple drop, we find
that the core in these two states enhance the deformation of the compound drop
and also affects its migration. The slow outward motion of the core in the
stable state pushes the compound drop slightly towards the wall, while the
revolving core in the limit-cycle state makes the compound drop to oscillate in
its equilibrium position. We also find that the direction of migration of the
compound drop affects the eccentricity of the core.

We also conducted a parametric study to understand the influence of each
parameter on the core-shell interaction. The key results are the following.

\begin{itemize}

\item Increasing any parameter results in the transition from limit-cycle state
  to stable state, which occurs due to the increased deformation of the core.

\item The deformation of the compound drop is enhanced only at intermediate core
  sizes. At extreme sizes, the deformation of the compound drop is nearly same
  as the simple drop. This behavior is due to either very less interaction at
  small core sizes or the increased interactions at large sizes.
  
\item The migration of the compound drop moves increasingly towards the wall
  with increasing core size.
  
\item Both the core and the shell become symmetric about the centerline only
  when the equilibrium position of the compound drop is at the center. Even in
  this symmetric position, the core is still eccentric where it occupies the
  front of the shell.

\end{itemize}

The similarities with the results reported in the literature shows that the
essential physics of the compound drops in Poiseuille flows is captured
sufficiently by our 2D simulations. We believe these results provide a better
understanding of the compound drops in pressure-driven flows, which can be used
to identify favorable flow conditions for various applications.

\section{Supplementary material}
The supplementary material contains equilibrium data for all the cases
simulated, and plots showing the effect of initial conditions on the dynamics of
the core and the compound drop.

\begin{acknowledgments}
  \textbf{V.T.G} would like to acknowledge the financial support from IIT Madras
  via the institute post-doctoral fellowship. \textbf{V.T.G} is grateful for
  insightful discussions with Mr. Akash Choudary, Mr. K.V.S Chaithanya, and
  Mr. Prathamesh M. Vinze. We also thank the anonymous referees for providing
  valuable suggestions to improve the manuscript.
\end{acknowledgments}

\section{Data Availability}
  The data that supports the findings of this study are available within the
  article and its supplementary material.

\appendix
\section{Derivation of the Source term $S_p$}
\label{app_01}
In our simulations, we have used periodic boundary conditions to reduce the
computational resources by limiting the size of the computational domain. The
use of periodic boundary condition requires the pressure gradient $(dp/dx)$
along the axial direction which drives the flow, to be included as a source
term $\mathbf{S_P}$ in Eq. (\ref{eq_mom}). This pressure gradient for plane
Poiseuille flow can be expressed in terms of the velocity profile as shown
below.

The velocity profile for the plane Poiseuille flow is given as,
\begin{equation}
  U(y) = \left(\frac{1}{2\mu_c}\right)\left(\frac{dp}{dx}\right)(Hy-y^2).
\end{equation}

The maximum velocity $U_m$, which occurs at the center of the channel ($y=H/2$)
is,
\begin{equation}
  U_m = \left(\frac{H^2}{8\mu_c}\right)\left(\frac{dp}{dx}\right).
\end{equation}
Rewriting the above equation in terms of pressure gradient we get the source term,\begin{equation}
  \mathbf{S_P} = \frac{8\mu_cU_m}{H^2}\hat{x}, 
\end{equation}
where $\mathbf{S_p}$ is the pressure gradient.
\section{Grid independence study}
\label{app_1}
\begin{figure*}[t]
  \subfloat[]{
    \centering
    \includegraphics[scale=0.7]{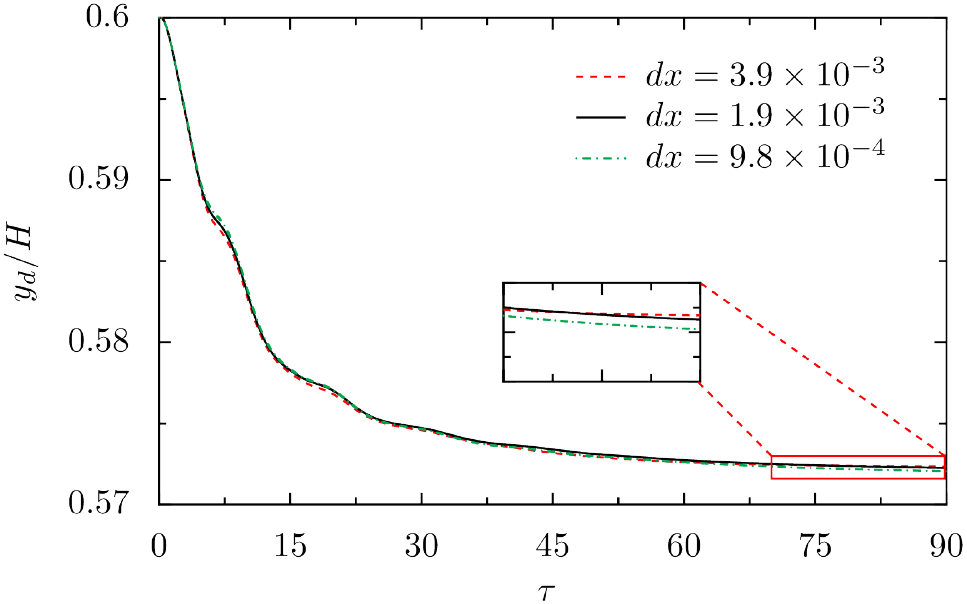}}
    \subfloat[]{
    \centering
    \includegraphics[scale=0.7]{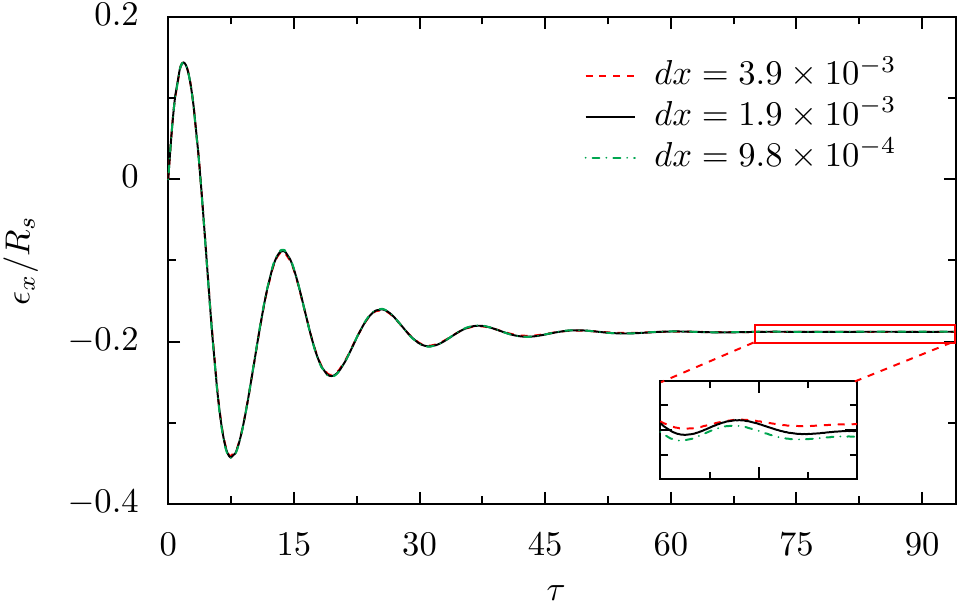}}
  
\caption{Effect of mesh resolution on (a) trajectory of the compound drop and
    (b) eccentricity of the core. The parameters used for this simulation are
    $\lambda=0.2$, $K=0.5$, $\mu_r=1$ and $Ca=1$.}
  \label{fig_grid}
\end{figure*}

We perform a grid convergence test to ensure that the results are independent of
grid resolution. Since we use adaptive mesh refinement, we simply vary the
maximum refinement levels between 8 to 10. The size of the smallest region
resolved for levels 8, 9 and 10 are $3.9 \times 10^{-3},~ 1.9 \times 10^{-3}$
and $9.8 \times 10^{-4}$, respectively. The results of the test are shown in
Fig. \ref{fig_grid} for the trajectory of the compound drop and the eccentricity
of the core. It can be concluded that the variation in the measured parameters
between different mesh sizes is minimal, so we use a maximum refinement of 9
levels for all our simulations.

\section{Validation}
\label{app_2}
\begin{figure*}[t]
  \subfloat[]{
    \centering
    \includegraphics[height=0.14\textwidth]{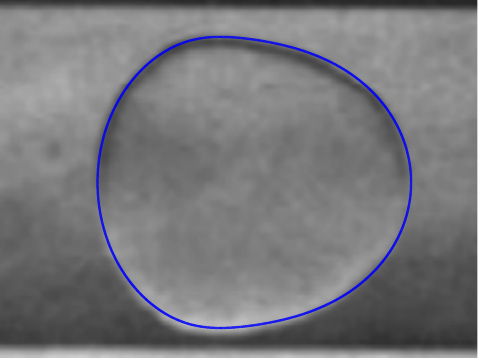}
  }
\subfloat[]{
    \centering
    \includegraphics[height=0.14\textwidth]{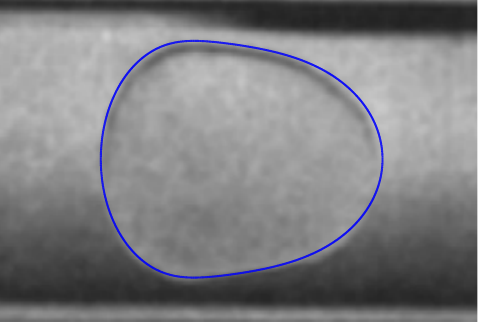}
  }
\subfloat[]{
    \centering
    \includegraphics[height=0.14\textwidth]{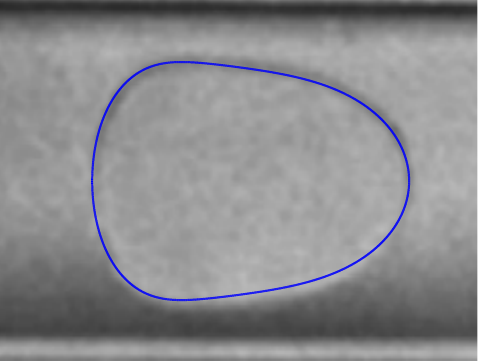}
  }
  
  \caption{Comparison of steady state drop shapes between the experiments
    \cite{Olbricht1992} and the simulations (shown in blue lines) at  (a)
    $Ca=0.05$ (b) $Ca=0.1$ (c) $Ca=0.16$. Other parameters are $\mu_r=1.05$ and $\lambda=0.95$. }
  \label{fig_valid_axi}
\end{figure*}

We verify the validity of the numerical solver used in this work with the
experimental and numerical results reported in the literature
\cite{Mortazavi2000,Olbricht1992} on simple drops in Poiseuille flow. First we
validate the solver for drops migrating in a capillary tube \cite{Olbricht1992}
at different capillary numbers. This problem is simulated under Stokes regime in
axisymmetric configuration, and the computational setup is a rectangular domain
of size $6 \times 1$. The steady state shapes of the drop from the simulations
are compared with the shapes from the experiments in Fig. \ref{fig_valid_axi},
and we find that the  solver is able to predict the shapes of the drop
with sufficient accuracy.

\begin{figure*}[t]
  \subfloat[]{
    \centering
    \includegraphics[scale=0.7]{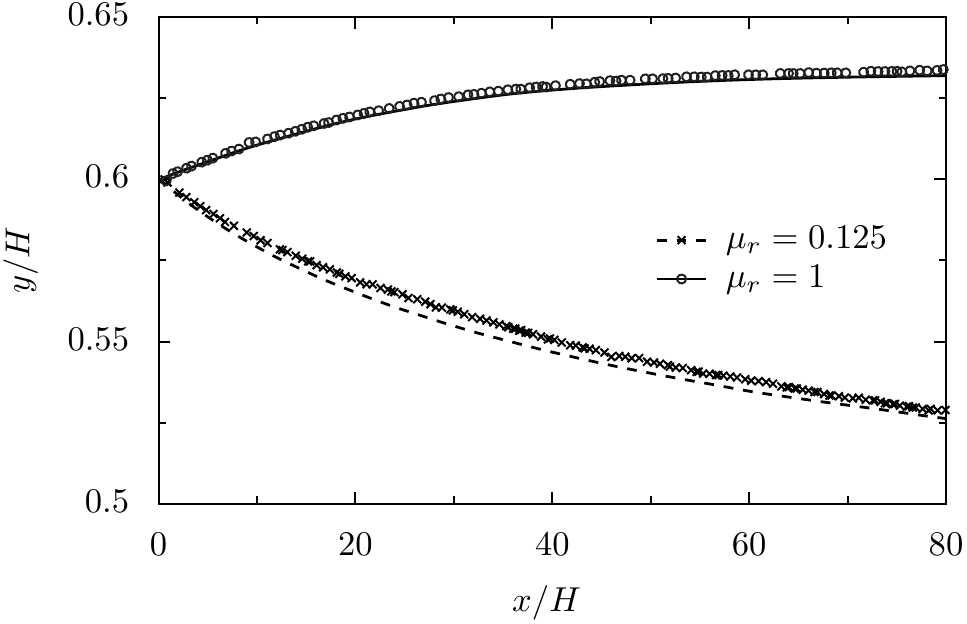}
  }
  \caption{Trajectory of the drop at
$Re(=U_mH/\nu_c)=1,~Ca=0.25,~\lambda=0.125$. The lines represent current
simulation and the markers represent the simulation data from
\cite{Mortazavi2000}.}
  \label{fig_valid_mig}
\end{figure*}

In the second case, we verify the solver for the migration of a small drop
($\lambda=0.125$) in a plane Poiseuille flow \cite{Mortazavi2000} at different
viscosity ratios. We solve the full Navier-Stokes equations here (including the
inertial terms) in a square computational domain of size
one. Fig.\ref{fig_valid_mig} compares the trajectory of the compound drop
obtained from our simulations with the results in the literature. It shows a
good agreement with the reported results from the literature.

\nocite{*}
\bibliography{References}

\end{document}